\documentclass[3p,times,authoryear]{elsarticle}
\PassOptionsToPackage{dvipdfmx}{graphicx}
\usepackage{amsmath}
\usepackage{amsthm}
\usepackage{comment}
\usepackage{amssymb}
\usepackage{comment}
\usepackage{graphicx}[dvipdfmx]  
\usepackage[utf8]{inputenc}
\usepackage[capitalise]{cleveref}
\usepackage{xcolor}
\usepackage[subrefformat=parens]{subcaption}
\usepackage{url}

\usepackage{nccmath} 

\usepackage{comment}

\biboptions{comma,sort&compress}

\everymath{\displaystyle}

\begin{document}

\begin{frontmatter}
\renewcommand{\thefootnote}{\fnsymbol{footnote}}
\title{A hybrid neural network for real-time OD demand calibration under disruptions} 
\address[monash]{Monash Institute of Transport Studies, Monash University, Australia}
\address[Seoul]{Department of Transportation Engineering, the University of Seoul, Korea}

\author[monash]{Takao Dantsuji\corref{cor1}}\ead{Takao.Dantsuji@monash.edu} \cortext[cor1]{Corresponding authors.}
\author[monash]{Dong Ngoduy\corref{cor2}}\ead{Dong.Ngoduy@monash.edu} \cortext[cor2]{Corresponding authors.}
\author[monash]{Ziyuan Pu}
\author[Seoul]{Seunghyeon Lee}
\author[monash]{Hai L. Vu}
\begin{abstract}

Existing automated urban traffic management systems, designed to mitigate traffic congestion and reduce emissions in real time, face significant challenges in effectively adapting to rapidly evolving conditions. Predominantly reactive, these systems typically respond to incidents such as accidents or extreme weather only after they have transpired. A promising solution lies in implementing real-time traffic simulation models capable of accurately modelling environmental changes, thereby minimising the adverse effects of disruptions on traffic flow and driver experience. Central to these real-time traffic simulations are origin-destination (OD) demand matrices, which serve as critical parameters. However, the inherent variability, stochasticity, and unpredictability of traffic demand complicate the precise calibration of these matrices in the face of disruptions. In response to this challenge, this paper introduces an innovative hybrid neural network (NN) architecture specifically designed for real-time OD demand calibration to enhance traffic simulations' accuracy and reliability under both recurrent and non-recurrent traffic conditions. The proposed hybrid NN predicts the OD demand to reconcile the discrepancies between actual and simulated traffic patterns. To facilitate real-time updating of the internal parameters of the NN, we develop a metamodel-based backpropagation method by integrating data from real-world traffic systems and simulated environments. This ensures precise predictions of the OD demand even in the case of abnormal or unpredictable traffic patterns.
Furthermore, we incorporate offline pre-training of the NN using the metamodel to improve computational efficiency. Validation through a toy network and a Tokyo expressway corridor case study illustrates the model's ability to dynamically adjust to shifting traffic patterns across various disruption scenarios. Our findings underscore the potential of advanced machine learning techniques in developing proactive traffic management strategies, offering substantial improvements over traditional reactive systems.
\end{abstract}

\begin{keyword}
real-time traffic simulation;  OD demand calibration; on-line learning; neural network; metamodel-based backpropagation
\end{keyword} 
\end{frontmatter}
\renewcommand{\thefootnote}{\arabic{footnote}}
\section{Introduction} 
\subsection{Background}
About 30\% of urban traffic congestion comes from non-recurrent traffic congestion and delays due to disruptions such as incidents, accidents and adverse weather. Managing non-recurrent traffic congestion is challenging, while our understanding of the behaviour of road users in such a fast-changing environment is limited and often based on incomplete, untimely information \citep{khattak2010spatial, zhong2017forecasting}.  This limits the potential of any current urban traffic management systems to deal with disruptions effectively. Traditional management systems, largely
designed for recurrent and predictable traffic patterns, are often reactive, such as feedback controls \cite[e.g.,][]{papageorgiou1991alinea}. The systems can respond to events only after they have occurred. This reactive approach could delay response times and inefficiencies in coping with sudden changes in traffic conditions. A more proactive traffic management strategy is required to respond to anticipate and mitigate the impacts of non-recurrent traffic conditions. 

Real-time traffic simulation emerges as an essential tool in addressing the challenge of managing non-recurrent traffic conditions. Specifically, microscopic traffic simulators are the most detailed simulation tools that can describe traffic phenomena at the individual vehicle level. These can be used to assess the impacts of various traffic control strategies, including system-level control (e.g. signal control) and individual vehicle control (e.g. autonomous vehicle control). By leveraging real-time data collected from limited traffic measurement points and continuous parameter updates, these simulations provide accurate short-term traffic conditions \cite[e.g.,][]{zhou2007structural}. This immediate prediction is crucial for developing proactive strategies that anticipate potential traffic bottlenecks rather than simply reacting to them after they occur. Furthermore, real-time simulations can test various traffic management strategies at a low cost in a virtual environment before implementing them in the real world, thereby enhancing the decision-making process under uncertain conditions. Consequently, real-time traffic simulation can predict short-term traffic conditions under unpredictable disruptions and proactively improve overall traffic efficiency and safety. 

Calibrating origin-destination (OD) demand matrices, which are critical input (demand) parameters in traffic simulations, is complicated under non-recurrent traffic conditions. Each cell in an OD matrix represents the number of trips from one specific area or point to another, forming the backbone of traffic simulations. However, the inherent variability, stochasticity, and unpredictability complicate the accurate calibration of these matrices. Traditional calibration methods relying on historical traffic data \cite[e.g.,][]{dantsuji2022novel} may not capture the sudden changes in driver or route choices caused by such disruptions. The OD matrices must be updated frequently to reflect the current traffic state accurately. This need for real-time data integration and calibration demands advanced methodologies for OD calibration problems. 

Emerging artificial intelligence (AI) technology has made recent breakthroughs in predicting short-term traffic conditions and has the potential to trigger a shift from reactive to proactive traffic control.
For example, recently, deep learning (DL) approaches have been successfully applied to the transportation field problems such as traffic state estimation \cite[e.g., ][]{xu2020ge, shi2021physics, Wang2024}, traffic signal control \cite[e.g.,][]{chu2019multi, yu2023decentralized} and model parameter estimation \cite[e.g.,][]{mo2021physics}. While DL approaches have demonstrated robust capabilities in various transportation applications, they perform exceptionally well in environments with recurrent, predictable patterns where large volumes of historical data can be used for training. They neglect the fast-changing traffic patterns during local disruptions (e.g., spillages; localised extreme weather) and lead to biased traffic management decisions uninformed by the current conditions. Their predictions in non-recurrent traffic conditions can be biased due to uninformed current conditions. These conditions necessitate updating the machine learning model parameters for the OD calibration problems to reflect real-time changes in traffic dynamics accurately.

\begin{figure}[t]
\centering
\includegraphics[width=0.7\textwidth]{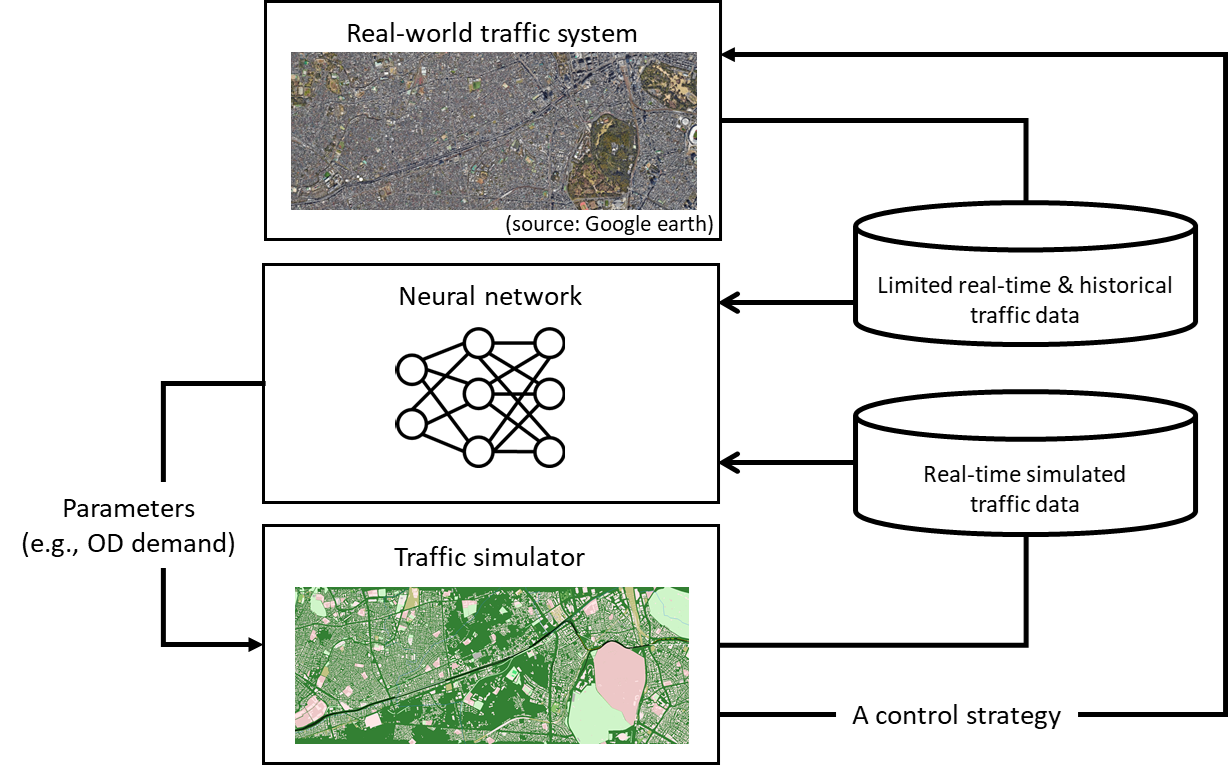}
\caption{A hybrid neural network for real-time simulation}
\label{fig:HNN}
\end{figure}

This paper proposes a hybrid neural network (NN) architecture for real-time OD demand calibration, as illustrated in Fig.~\ref{fig:HNN}\footnote{In this study, OD demand is considered as the parameters predicted from the NN. Other parameters, such as car-following parameters, are assumed to be calibrated offline by the existing methods \cite[e.g.,][]{ngoduy2012calibration, bai2021calibration}.}. The distinctive feature of this framework is its ability to seamlessly integrate traffic data from both (limited) real-time data and rich simulated environments, enhancing the accuracy and reliability of traffic predictions. This hybrid model continually updates the NN parameters to minimise discrepancies between actual and simulated traffic conditions at the observed locations. More specifically, to enable the update of the NN parameters via a black-box traffic simulator in real time, we propose a metamodel-based backpropagation method employing a problem-specific tractable metamodel. Updating the parameters of the trained NN via a metamodel-based backpropagation algorithm will yield an updated OD matrix as an output of the NN, which in turn serves as an (updated parameter) input into the simulation when the traffic changes. Improvements in the NN's predictions of traffic demands lead to enhanced precision in the simulated outputs, particularly in terms of spatio-temporal traffic conditions. By maintaining both the neural network and the traffic simulation models in a state of concurrent, real-time iteration, the framework adeptly accommodates dynamic traffic patterns that emerge from unpredictable disruptions.

\subsection{Related literature review}
\subsubsection{OD demand calibration}
Microscopic traffic simulators can arguably describe the most detailed interaction between vehicles, infrastructures, and control. Since well-calibrated simulators can reproduce reality, they are conventionally used for traffic planning, design, and operation tasks in an offline manner, including optimization for pricing \citep{chen2016time, dantsuji2021simulation, osorio2021efficient, zheng2022time} and network designs \citep{mesbah2011optimization, zheng2017macroscopic, li2022simulation}.  One of the critical parameters of traffic simulations is the origin-destination (OD) demand matrix wherein each cell represents the number of trips from a certain area (or point) to another. There is a vast literature on the {\it offline} OD demand calibration using multiple traffic data sources, such as loop detector \cite[e.g.,][]{toledo2004calibration, osorio2019dynamic, zhang2021improving}, probe data \citep{toledo2003calibration}, and point-to-point travel time from Bluetooth \citep{barcelo2013kalman, dantsuji2022novel}. Utilizing historical data, offline OD calibration approaches have been proposed mainly based on the general purpose algorithms such as Stochastic Perturbation Simultaneous Approximation (SPSA)-based \cite[e.g.,][]{balakrishna2007calibration, vaze2009calibration, cipriani2011gradient, ben2012dynamic, cantelmo2014adaptive, tympakianaki2015c, antoniou2015w}, Genetic Algorithm based \cite[e.g.,][]{stathopoulos2004hybrid} approaches. To tackle the computational complexity, computational efficient approaches such as metamodel-based optimization \cite[e.g.,][]{osorio2019dynamic,OSORIO201918,patwary2021metamodel} have also been proposed. 

In real-time traffic simulation applications, the offline OD calibration problems were extended to online OD calibration problems.  Similar to the offline OD calibration problems, the general purpose algorithms are mainstream of online OD calibration owing to their applicability to a wide range of problems. For example, Kalman filter-based approaches \citep{okutani1984dynamic,ashok2000alternative,zhou2007structural, marzano2018kalman, zhang2021improving}.  As indicated in \cite{zhang2021improving}, the accuracy under congestion may be low due to violations of the Markovian assumption in online applications of the extended Kalman filter. Real-time implementation of traffic control and incidents may also violate the Markovian assumption. Furthermore, although these algorithms can be applied to a wide range of problems, computational efficiency is not a priority. 

\subsubsection{Machine learning/AI in transportation}

With the recent development of AI technology, research to solve complex and large-scale problems that exist in reality is becoming more feasible. The use of AI technology is prominent in various fields (e.g., autonomous vehicles, healthcare, and engineering) and has the advantage of being able to derive and learn the relationship between factors inherent in complex problems that are difficult to explain. This improves our ability to generate Data Driven Models (DDMs) that can extract meaningful and actionable information from data. For example, Deep, Multi-Task, Transfer, and Semi-Supervised learning algorithms, together with rigorous statistical inference procedures, make it possible to transform large and heterogeneous volumes of distributed and hard-to-interpret pieces of data into meaningful, easy-to-interpret information. Some initial efforts have been made to apply DDMs for predicting and easing traffic congestion by training models with massive data and coming up with relevant traffic signal timings \citep{wang2019enhancing}. Unlike existing tools, including Google Maps or in-vehicle navigators, DDMs provide the potential to understand traffic patterns with real-time data to ensure accurate and reliable predictions across a whole network. 

Specifically deep learning (DL) approaches are powerful tools to predict patterns behind data and have gained popularity in recent years. DL approaches were successfully applied to the transportation field problems such as causal effects for the crashes \cite[e.g., ][]{li2024}, traffic state estimation \cite[e.g., ][]{xu2020ge, shi2021physics, Wang2024,jiang2023a,jiang2023b}, traffic signal control \cite[e.g.,][]{chu2019multi, yu2023decentralized,lee2022,lee2019b,jiang2021}, model parameter estimation \cite[e.g.,][]{mo2021physics,lee2019a}.  Traffic demand estimation\footnote{It is important to note that demand estimation is to estimate general OD matrices for traffic planning and design \cite[e.g.,][]{behara2020novel}, and demand calibration is to calibrate OD matrices for stochastic traffic simulators. } is also one of the most well-studies DL applications, such as not only vehicular demand estimation \cite[e.g., ][]{krishnakumari2020data, khalesian2024improving, owais2024deep}, but also for-hire-vehicles demand estimation \cite[e.g., ][]{liu2019contextualized, qian2020short, ke2021predicting, zhang2021dneat}, and rail passenger demand \cite[e.g., ][]{zhang2021short}. 

Although DDMs scale well to the amount of data available in a large-scale network, they are not effective at deducing disruptions. Furthermore, DDMs are commonly utilised as black boxes whose inputs and outputs interrelation is not visible to the users. This means that understanding how their predictions are obtained and characterising confidence intervals, tends to be challenging, especially in case of disruptions, where the traffic patterns change significantly from the historical data and are hard to predict. The development of proactive, real-time traffic control strategies dealing with disruptions in the network thus requires a distinct, ML-based approach that can predict the ever-changing traffic environment where the previously learnt models are not accurate. However, it is unrealistic and impracticable to re-train the models with such continuously changing sequences of real-time data. For example, disruptions due to incidents or adverse weather conditions lead to significant changes in traffic flows in a relatively short term. Therefore, unusual or non-recurring situations cannot be inferred accurately from existing ML-based approaches. In contrast, the cause of non-recurrent disruptions can be modeled accurately by traffic flow models after detection using a proper method \cite[e.g.,][]{li2022real}. This led to a hybrid method that employs both data-driven and model-based traffic predictions proposed in literature \cite[e.g, ][]{huang2020physics, shi2021physics, kim2024hybrid}.  However, little research exists on OD calibration problems. This could be because of the lack of a methodology that connects DDM for OD calibration and a black-box traffic simulator to enable a two-way flow of information between them. 

\subsubsection{Major contributions}

As summarized in Table.~\ref{tab:model_comparison}, model-based approaches leverage domain knowledge (traffic models) to provide structured and interpretable  predictions, but they often struggle with scalability and face challenges in real-time parameter calibration (e.g.,  OD demand) as discussed above. Conversely, data-driven approaches, while scalable and capable of capturing complex patterns from vast datasets, may require extensive re-traininig  to accurately respond to rapidly changing traffic condition under disruptions, making real-time adjustment difficult. 

\begin{table}[t!]
\centering
\caption{Comparison of Model-Based and Data-Driven Approaches}
\label{tab:model_comparison}
\begin{tabular}{|l|c|c|}
\hline
& Model-Based Approach  &   Data-Driven Approach    \\ \hline
\textbf{Model Type}     &   Domain knowledge (traffic models)     &  Data-driven (black-box) models  \\ 
& & (machine learning algorithms) \\ \hline
\textbf{Representation} &   Causal effect relationship       & Associational relationship  \\
& between variables & between variables\\ \hline
\textbf{Structure}      &   Domain knowledge is required:                                      & No domain knowledge is required:  \\
& dynamic mapping of input to output & static mapping of input to output \\ \hline
\textbf{Validation}  & Parameter calibration                  &  Environment remains unchanged    \\
&& before and after training \\ \hline
\textbf{Scalability}             & Limited  & Scalable               \\ \hline
\textbf{Impacts of}     &  Predictable   &  Unpredictable   \\
\textbf{disruptions} &&\\ \hline
\end{tabular}
\end{table}

To  bridge the gap between model-based and data-driven approaches to enhance traffic condition prediction accuracy under non-recurrent traffic conditions, this paper aims to implement a hybrid ML framework which utilizes a traffic simulation model as a digital twin of the network to provide an alternative to DDMs, thereby improving the robustness of traffic flow behaviour according to the network's physical constraints. More specifically, the input (i.e. OD flows) is entered into the traffic model to determine the spatiotemporal traffic conditions along the link including any changes detected in the network via other means \cite[e.g.,][]{li2022real} based on the real-time observed traffic data.  As a result, it can better predict how traffic patterns are related to disruptions which might happen anywhere inside the link at any time. This computational output from the traffic model is an intermediate input to the DDMs. 

The spatiotemporal traffic conditions generated by the traffic model will provide rich additional synthetic real-time data (together with the actual data from traffic sensors) as input to the continuous learning of the DDM, which consequently enhances its output accuracy in case of disruptions. Moreover, this approach utilises a DDM to estimate the input (i.e., the OD demand), and the DDM output will be fed back to the traffic model in the next iteration. This iterative process will further improve both the traffic model (or the network digital twin) and the DDM as it operates.  To this end, the proposed approach allows for the effective and scalable modelling of complex systems based on the physical knowledge about them and deducing meaningful information about the systems by simulating the traffic model. Achieving this aim will open up significantly the capabilities of urban traffic control and management to deal with disruptions as it will support a holistic view of the controlled traffic region and increase response flexibility and effectiveness.

\vspace{10pt}

The remainder of this paper is organized as follows. Section 2 shows the development of the framework. Section 3, we validate the proposed approach with a toy network and show numerical results with the case study of Tokyo expressway corridor. Finally, we conclude this paper in Section 4.

\section{Methodology}
\subsection{Problem setting of OD calibration}
We first introduce the problem setting of OD calibration for real-time simulation. Consider a corridor or a network with multiple origins and destinations. Traffic measurements at location $k$ (where $k$ is a point of the set $K$ of the measurement points) can be accessed in real-time. The goal is to predict the traffic demand at the next time iteration $t+1$, to minimize the distance between the actual and simulated traffic measurements at $t+1$. The prediction problem is formulated as follows: 
\begin{align} \label{eq:obj_function}
    \min_{\bf D_{t+1}} \frac{1}{|K|}\sum_{k \in k} \left( \hat{f}_{t+1}^{k} - \mathbb{E} \left[ f^{k}(t+1 ;  {\bf D_{t+1}})  \right]  \right)^2 + \delta  \frac{1}{|I|}\frac{1}{|J|}\sum_{i \in I} \sum_{j \in J} \left(  D^{ij}_{t+1} - \overline{D}^{ij}_{t+1} \right)^2,
\end{align}
where ${\bf D_{t+1}}$ is the vector of the traffic demand at next time iteration $t+1$, $\hat{f}_{t+1}^k$ is the actual traffic measurement at time $t+1$ at (observed) location $k$, $\mathbb{E} \left[ f^{k}(t+1 ;  {\bf D_{t+1}})  \right]$ is the expected traffic measurement from the stochastic traffic simulation at time $t+1$ at location $k$ with demand $ {\bf D_{t+1}}$. $\overline{D}^{ij}_{t+1}$ is the a priori demand from origin $i$ to destination $j$ at $t+1$, which can be estimated using existing OD demand estimation methods with historical data, $I$ and $J$ are the set of the origins and destinations, respectively, and $\delta$ is the weight parameter representing the importance of a priori demand.  

The first term of equation (\ref{eq:obj_function}) denotes the gap between the actual and simulated traffic states. Common traffic measurements include traffic counts, speed, and density. However, due to the U-inversed shape of the flow-density fundamental diagram, traffic count data is not suitable for oversaturated conditions as it does not distinguish between uncongested and congested traffic states. This paper uses traffic density as traffic measurement ${\bf \hat{f}_{t}}$. The second term is the error between predicted and a priori demand. If the available historical data is of poor quality for estimating a priori demand, or if demand changes significantly due to traffic control or incidents, the weight parameter should be set to be low value. Conversely, setting the value to 1 implies equal importance for both terms in equation (\ref{eq:obj_function}).

The traffic demand prediction challenge arises from the complexity, stochasticity, and uncertainty inherent in urban transportation. The proposed real-time calibration framework aims to find an effective predictor $ \Psi ( {\bf T_{t}, \hat{g}_{t}})$, based on time-related variables and actual traffic measurements at the current time step: 
\begin{align}
 {\bf D_{t+1}} = \Psi ( {\bf T_{t}, \hat{g}_{t}}),
\end{align}
where $\bf T_t$  is a vector of time-related variables (e.g., hour, day-of-the-week) at the current step $t$ and  $\bf \hat{g}_t$ is traffic states at time $t$

\subsection{Proposed online learning framework}

A challenge in implementing the model in Fig.1 is that an algorithm that is suitable at the current point in time may become inappropriate when the characteristics of traffic flow change sharply due to network disruptions as well as the impact of the adaptive signal control strategies. To improve the predictions of the NN model, we will utilise knowledge from off-line estimation results on multiple days to represent a priori estimate of the regular traffic patterns. Reliable historical data will serve as an informative source under normal conditions (i.e., in a static environment). Then our proposed framework is able to map the prior estimate and real-time data from both predictive traffic simulation models and actual data to determine its performance. In essence, our proposed method is more advanced than the traditional ML algorithms due to its adaptation using predictive traffic simulation models to generate additional synthetic data in real-time. The goal is to iteratively adjust the previously learned algorithms in the original environment (i.e., the a priori estimate) whenever the data pattern changes using synthetic data obtained from simulation models with the disruptions detected by the real-time measurements.

\begin{figure}[t!]
\centering
\includegraphics[width=\textwidth]{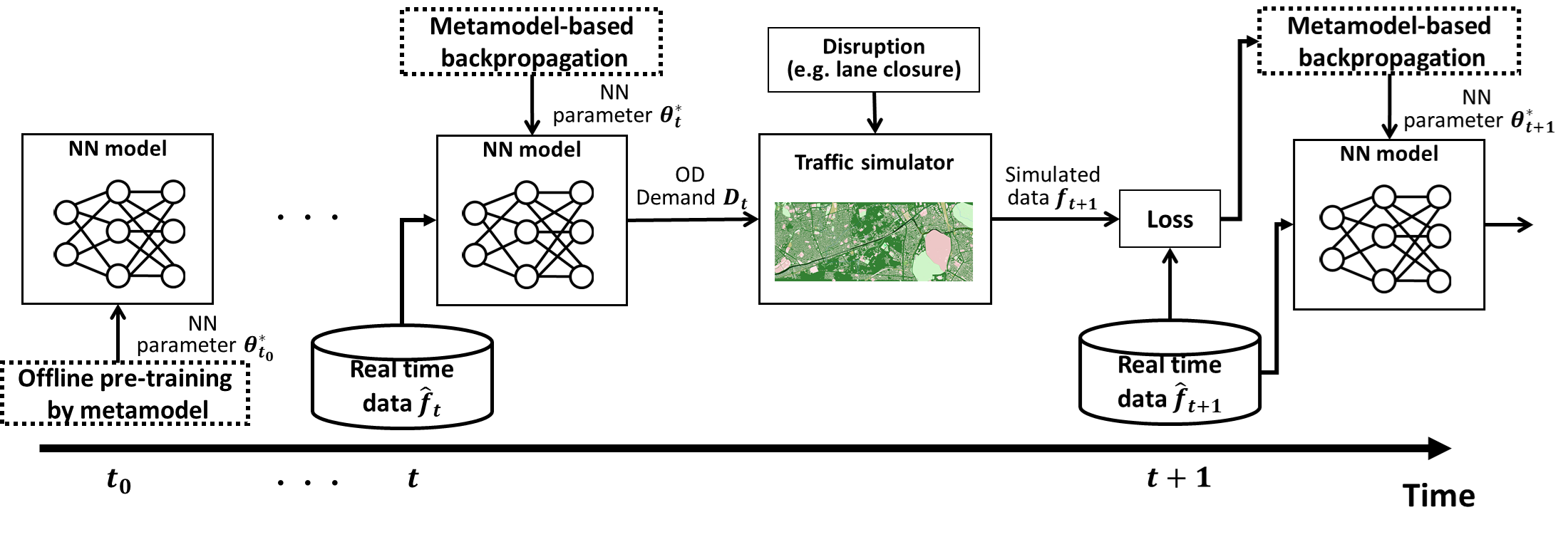}
\caption{The proposed real-time simulation framework}
\label{fig:flowchart}
\end{figure}

We employ deep neural network (DNN) for the traffic demand predictor, denoted as $\Psi ( {\bf T_{t}, \hat{g}_{t}; \theta})$ parameterized by $\theta$.  The goal of the framework is to find optimal parameters $\theta^*_t$ that minimize the loss function defined as
\begin{align}
\theta_{t}^* = \underset{{\Psi ( {\bf D_{t}, \hat{g}_{t}; \theta_t}) }}{\arg \min} 
 \mathcal{L}_f + \delta \mathcal{L}_d,
\end{align}
where $\mathcal{L}_f =  \frac{1}{|K|}\sum_{k \in K} \left( \hat{f}_{t+1}^{k} - \mathbb{E} \left[ f^{k}(t+1 ;  \Psi ( {\bf D_{t}, \hat{g}_{t}; \theta_t)}  )  \right]  \right)^2$ and $\mathcal{L}_d=   \frac{1}{|I|}\frac{1}{|J|}\sum_{i \in I} \sum_{j \in J} \left( {\Psi^{ij} ( {\bf D_{t}, \hat{g}_{t}; \theta_t}) }  - \overline{D}^{ij}_{t+1} \right)^2$. $ {\Psi^{ij} ( {\bf D_{t}, \hat{g}_{t}; \theta_t}) } $ is the predicted demand from origin $i$ to destination $j$ at $t+1$ from the demand predictor ${\Psi ( {\bf D_{t}, \hat{g}_{t}; \theta_t}) }$. $ \mathcal{L}_f $ and $\mathcal{L}_d$ are the loss functions regarding the traffic measurement and the traffic demand, respectively.
To overcome the aforementioned challenges -- the real-time model parameter update of the DNN, the parameter update through a traffic simulator, and the computational load-- we propose a real-time simulation framework as depicted in Fig. \ref{fig:flowchart}.

Suppose that at time $t$, the agent learnt the optimal parameters $\theta_t^{*}$  with respect to the policy approximation (3). It is worth mentioning that in a static environment, the optimal parameters are non-time varying so $\theta^{*}$ can be estimated from (massive) historical data. When the environment changes at time $t+1$ due to disruptions, our goal is to adjust the existing knowledge $\theta_t^{*}$ to a new one $\theta_{t+1}^{*}$ that can achieve an optimum objective in the new environment defined in equation (3) with initialization of $\theta_{t+1}^{*} \leftarrow \theta_t^{*}$ . Continually, the agent iteratively adjusts the existing knowledge to a new one, $[\theta_{t+1}^{*},\theta_{t+2}^{*},\cdots]$  whenever the environment changes. The proposed method can thus rapidly adapt to environmental changes in real-time and continuously adjust the rest of the learning process.

In summary, the proposed framework in Fig. 2 consists of three main steps: (I) Offline pre-training by a metamodel, (II) real-time OD demand calibration, and (III) Metamodel-based backpropagation. By utilzing historical data, we train the DNN model offline through the metamodel which is analytically tractable and differentiable, instead of using a black-box traffic simulator. Because of the simplicity of the metamodel, this offline training is computationally efficient and provides a good starting point for $\theta$. Then, the pre-trained DNN is employed for real-time OD demand calibration. Subsequently, relying on the differentiability of the metamodel, we develop a metamodel-based backpropagation during online learning.

\subsection{Metamodel-based backpropagation}
\begin{figure}[t!]
\centering
\includegraphics[width=\textwidth]{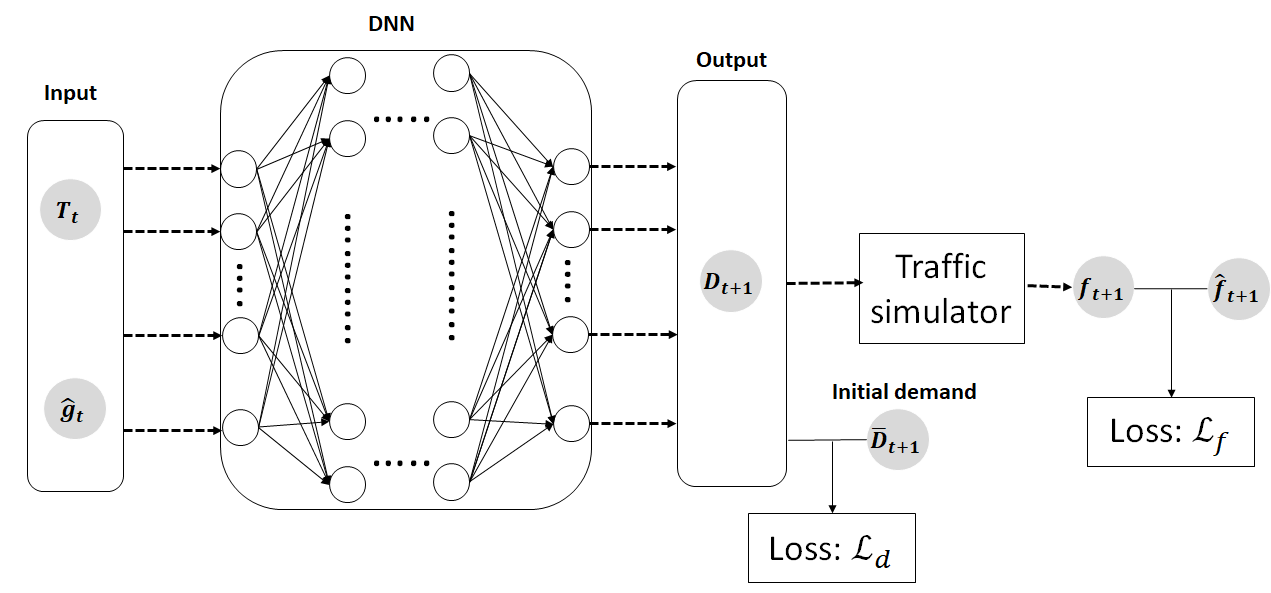}
\caption{The integration of DNN with traffic simulator}
\label{fig:DL_simulator}
\end{figure}

Predictions from the DNN's previous knowledge under traffic control or incident conditions may not be effective due to changes in traffic patterns caused by travel behaviors or road infrastructure changes. The DNN parameters must be updated in such scenarios. In the proposed framework, the DNN model is used for predicting OD demand $\bf D_{t+1}$ at the next time step $t+1$ from the input data such as traffic states $\bf \hat{g}_t$ and time-related variables $\bf T_t$ at the current step $t$, as depicted in Fig.~\ref{fig:DL_simulator}. Then, the predicted demand is used as the input of traffic simulators, and the output from the simulators (e.g., traffic states $\bf f_{t+1}$ at the next time step $t+1$) is used to compute the loss function $\mathcal{L}_f$ from the real values  $\bf \hat{f}_{t+1}$ to update the DNN parameters. However, as traffic simulators are normally black-box, the estimated loss cannot be propagated backwards for the DNN parameter update. It is thus challenging to update the DNN parameters online when integrating traffic simulators with DL approaches.

\begin{figure}[t!]
\centering
\includegraphics[width=\textwidth]{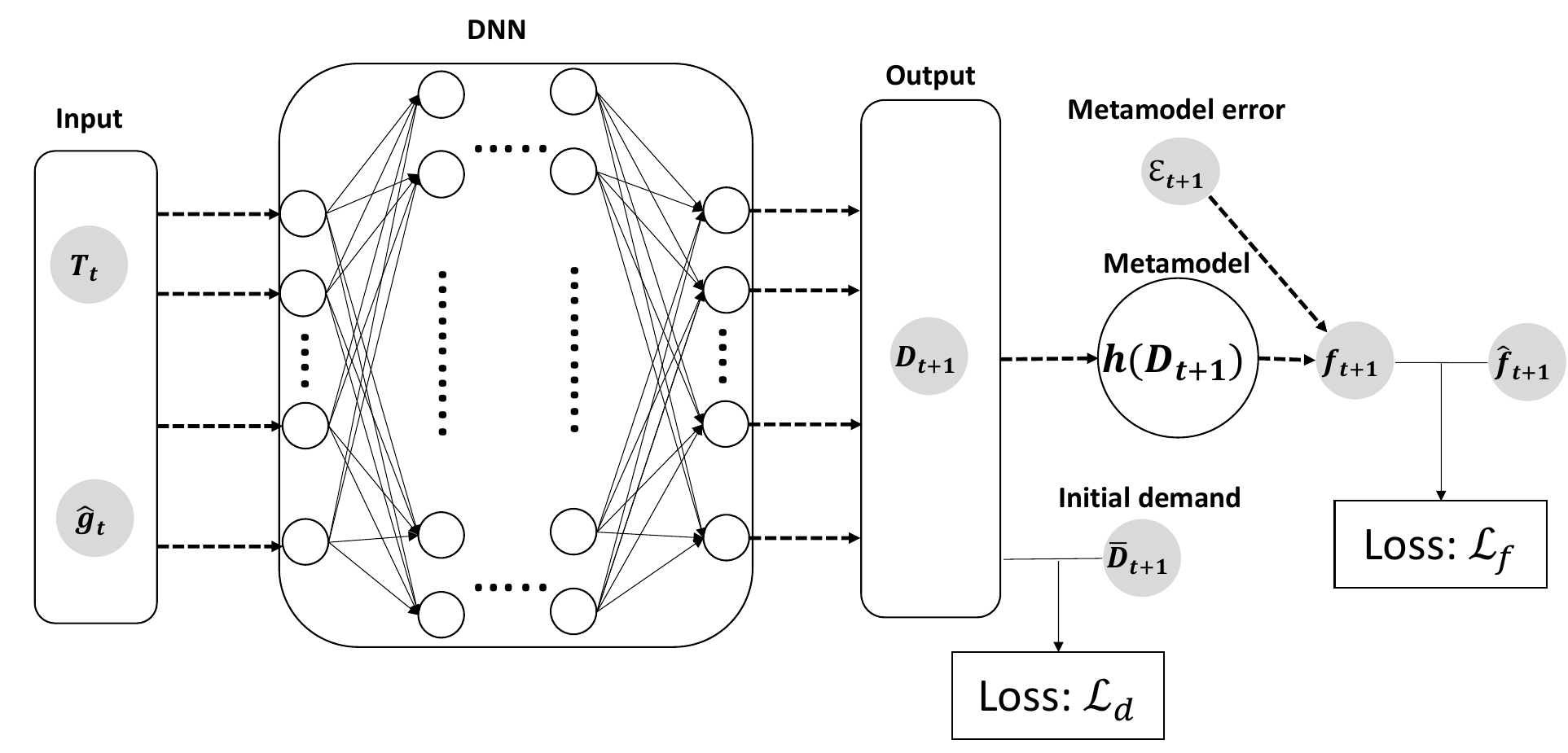}
\caption{The architecture of metamodel-based backpropagation}
\label{fig:metamodel_BP_architecture}
\end{figure} 

To address this challenge, we propose metamodel-based backpropagation, as depicted in Fig \ref{fig:metamodel_BP_architecture}. We substitute the traffic simulation model with the problem-specific metamodel $h({\bf D_{t+1}})$ when updating the parameters. The metamodel, denoted as $h({\bf D_{t}})$, is a static traffic assignment map that describes the relationship between the demand ${\bf D_{t}}$ and the traffic measurements ${\bf \check{f}_{t}}$. Given that the metamodel is analytically tractable, the estimated loss can be propagated backwards via the metamodel through the DNN, as depicted in Fig. 4. As the simulated traffic measurement ${\bf {f}_{t+1}}$ can be obtained from the simulator, the loss function $\mathcal{L}_f$ is calculated as follows: 
\begin{align}\label{eq:loss_f_sim}
\mathcal{L}_f =  \frac{1}{|K|}\sum_{k \in K} \left( \hat{f}_{t+1}^{k} - {f}_{t+1}^{k}  \right)^2.
\end{align}
$\bf {f}_{t+1}$ can be expressed as 
\begin{align}\label{eq:meta_error}
{\bf {f}_{t+1}} = h({\bf D_{t+1}}) + \varepsilon_{t+1} 
\end{align}
where $\bf \varepsilon_{t+1}$ is the metamodel error, the gap in the traffic measurements between the metamodel and the traffic simulator. The gradient of the loss with respect to the output of the DNN is computable through the metamodel:
\begin{align}
\frac{\partial \mathcal{L}_f}{\partial {\bf D_{t+1}} } =  \frac{\partial \mathcal{L}_f}{\partial {\bf f_{t+1}^{i}} }  \frac{\partial {\bf f_{t+1}^{i}}} {\partial h( {\bf D_{t+1}})} \frac{\partial h( {\bf D_{t+1}})}{\partial {\bf D_{t+1}}}. 
\end{align}
This gradient is propagated backwards through the DNN to update the parameter $\theta_t^*$. The updated parameter is transferred to predict the demand at the next time step $t+2$ based on the time-related variables $\bf T_{t+1}$  and the actual traffic measurements ${\bf \hat{g}_{t+1} }$ and so on, as described in the previous section.

\subsection{Offline pre-training by metamodel}
Learning the parameters of  DNN using traffic simulators, even offline, is not an easy task, given that traffic simulators are black-box and computationally inefficient. Therefore, we use a problem-specific metamodel for offline pre-training.  Simulated traffic measurement ${\bf \check{f}_{t+1}}$ can be derived by the metamodel $h({\bf D_{t}})$ based on the predicted demand ${\bf D_{t+1}}$, the traffic measurement loss function is calculated as follows:
\begin{align}\label{eq:loss_f}
\mathcal{L}_f =  \frac{1}{|K|}\sum_{k \in K} \left( \hat{f}_{t+1}^{k} - \check{f}_{t+1}^{k}  \right)^2.
\end{align}
The gradient of the loss with respect to the output of the DNN (i.e., ${\bf D_{t+1}}$) can be computed via the metamodel. 
\begin{align} \label{eq:gradient_computation}
\frac{\partial \mathcal{L}_f}{\partial {\bf D_{t+1}} } =  \frac{\partial \mathcal{L}_f}{\partial {\bf \check{f}_{t+1}^{i}} }  \frac{\partial {\bf \check{f}_{t+1}^{i}}} {\partial h( {\bf D_{t+1}})} \frac{\partial h( {\bf D_{t+1}})}{\partial {\bf D_{t+1}}}. 
\end{align}
The parameter $\theta$ is learnt using the historical data, and is then applied in the real-time simulation framework. Due to the tractability and simplicity of the metamodel, the computational complexity is much lower than that of a traffic simulator. Moreover, the learnt parameter $\theta$ serves as a good starting point for the real-time simulation, further contributing to computational efficiency. 

\section{Case studies}
Our case studies focus on two specific examples: 1) a toy network where the ground truth is known and 2) a real-life Tokyo network during the Olympic event in 2020. In the toy network, we create an infrastructure disruption (i.e. capacity drop due to an incident/accident). In contrast, the disruption in the real-life network happened due to a sharp change in the demand during the Olympic event.
\subsection{Toy network}
\subsubsection{Simulation set-up}
\begin{figure}[t]
\centering
\begin{subfigure}[b]{0.5\columnwidth}
    \centering
    \includegraphics[width=\textwidth]{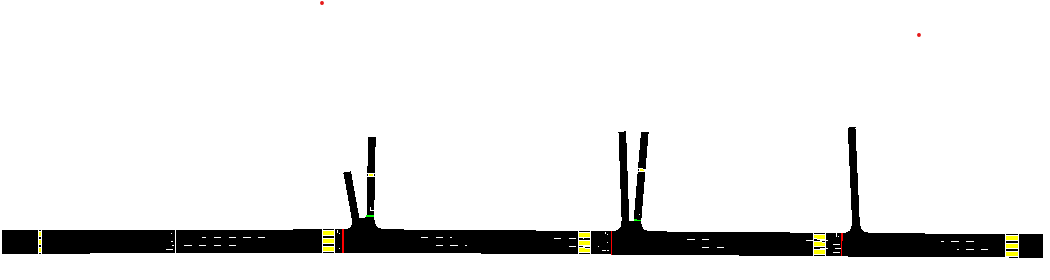}
    \caption{The toy network}
    \label{fig:toy_network}
\end{subfigure}%
\begin{subfigure}[b]{0.5\columnwidth}
    \centering
    \includegraphics[width=\textwidth]{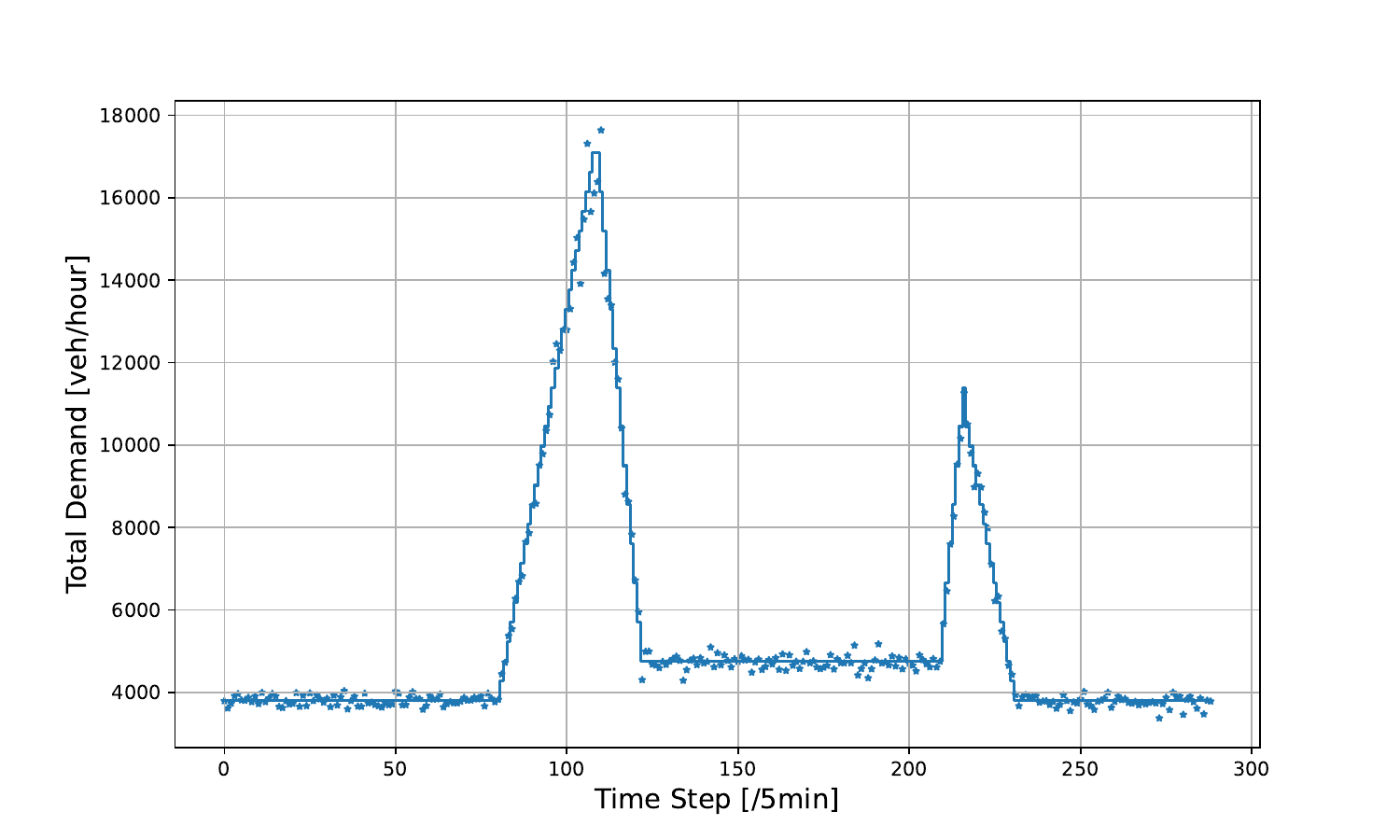}
    \caption{Demand profile}
    \label{fig:demand_profile}
\end{subfigure}
\caption{The toy network and its demand profile}
\label{fig:network_and_demand}
\end{figure}

We apply the proposed framework to a toy corridor network using the SUMO traffic simulator, as depicted in Fig. \ref{fig:toy_network}. This network contains 4 links equipped with loop detectors on a major road and 3 minor roads between these links. All links on the major road have three lanes, with link lengths ranging from 70 to 138 meters. The traffic signals operate under a multi-phase fixed-time schedule, with cycle lengths varying from 66 to 81 seconds. There are 9 OD pairs between the main road and minor roads. Traffic measurements are quantified and evaluated every 5 minutes. We assume a typical demand profile featuring morning and rush peak hours, as shown in Figure \ref{fig:demand_profile}. To generate true demand, we sample from a Gaussian distribution with a time-dependent mean and standard deviation set to 10 \% of the mean. An example is plotted in Fig. \ref{fig:demand_profile}.  We use ten days of simulation data as the pre-training dataset.

Since the case study site is a corridor network, the metamodel can be simply formulated as the model of \cite{osorio2019dynamic} as follows. 
\begin{align}
    k_k = c \frac{k_k^{jam}}{q_k^{cap}} \frac{\sum_{l \in L_k} D_l}{n_k}  \label{eq:density_estimation} 
\end{align}
where $k_k^{jam}$ is the jam density at measurement point $k$, $q_k^{cap}$ is the capacity of measurement point $k$, $n_k$ is the number of lanes at measurement point $k$, $D_l$ is the demand of OD pair $l$, $L_k$ is the set of the OD pairs of which path passes on measurement point $k$, and $c$ is the parameter. 

For the inputs of the DNN, we use the timestep as the time-related variables and both flow and speed as the traffic measurement variables. Note that the time-related variables are converted into the one-hot encoding data and traffic measurement variables are generalized.  The traffic measurement loss function is estimated by the error between simulated and actual traffic density. The time-dependent mean of traffic demand is used as a priori traffic demand. The DNN architecture is a fully connected feedforward neural network with 2 hidden layers and 48 hidden nodes in each hidden layer. The hyperbolic tangent function (tanh) is used as an activation function for each hidden neuron. We employ a stochastic gradient descent (SGD) algorithm with a learning rate of 1 to update the parameters.

\begin{figure}[t]
\centering
\includegraphics[width=0.5\textwidth]{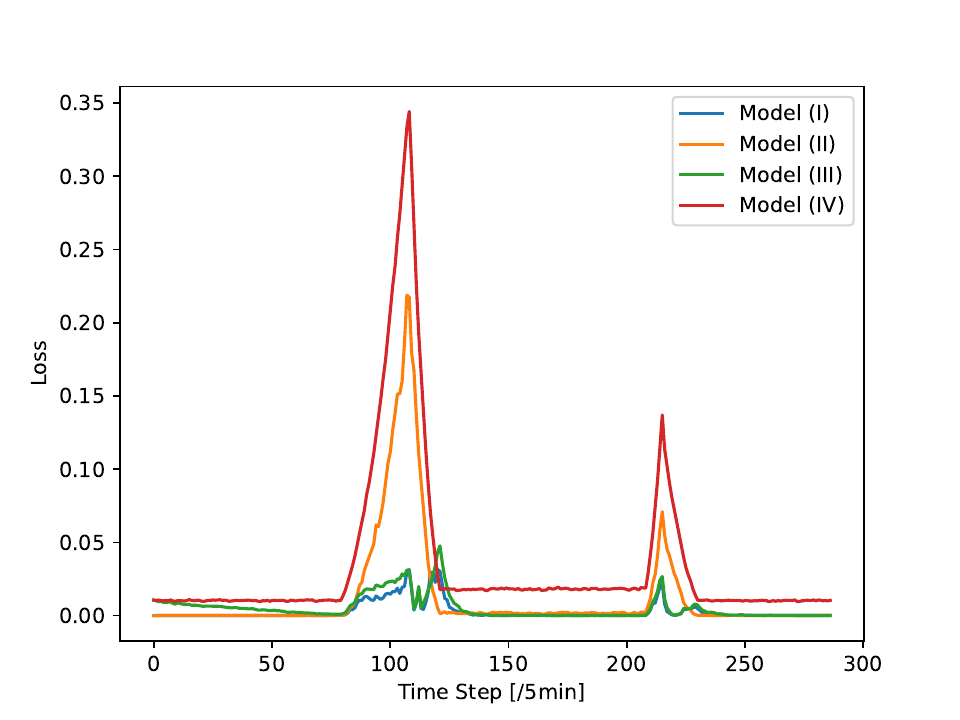}
\caption{Loss function over time steps}
\label{fig:loss_toy}
\end{figure}

\begin{table}[t]
    \centering
        \caption{Mean Squared Error (MSE) across the locations}
    \label{tab:mse_models}
    \begin{tabular}{cccccc}
        \textbf{Location} & \textbf{Model (I)} & \textbf{Model (II)} & \textbf{Model (III)} & \textbf{Model (IV)} \\
        \hline
        0 & 0.40 & 1.10 & 0.46 & 1.84 \\
        1 & 2.82 & 5.77 & 3.24 & 8.17 \\
        2 & 7.45 & 16.90 & 9.34 & 23.83 \\
        3 & 0.39 & 0.57 & 0.41 & 1.33 \\
        \hline
        &&&& [(veh/km)$^2$]
    \end{tabular}
\end{table}

\begin{figure}[p]
\centering
\begin{subfigure}[b]{0.5\columnwidth}
    \centering
    \includegraphics[width=\textwidth]{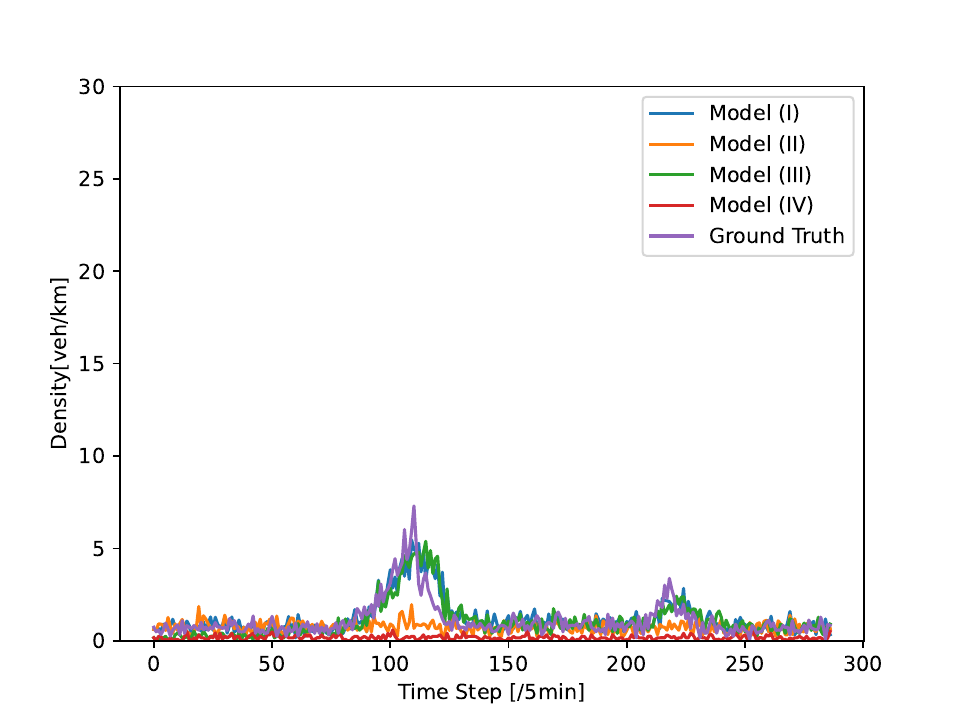}
    \caption{Location 0}
    \label{fig:Density_location_0}
\end{subfigure}%
\begin{subfigure}[b]{0.5\columnwidth}
    \centering
    \includegraphics[width=\textwidth]{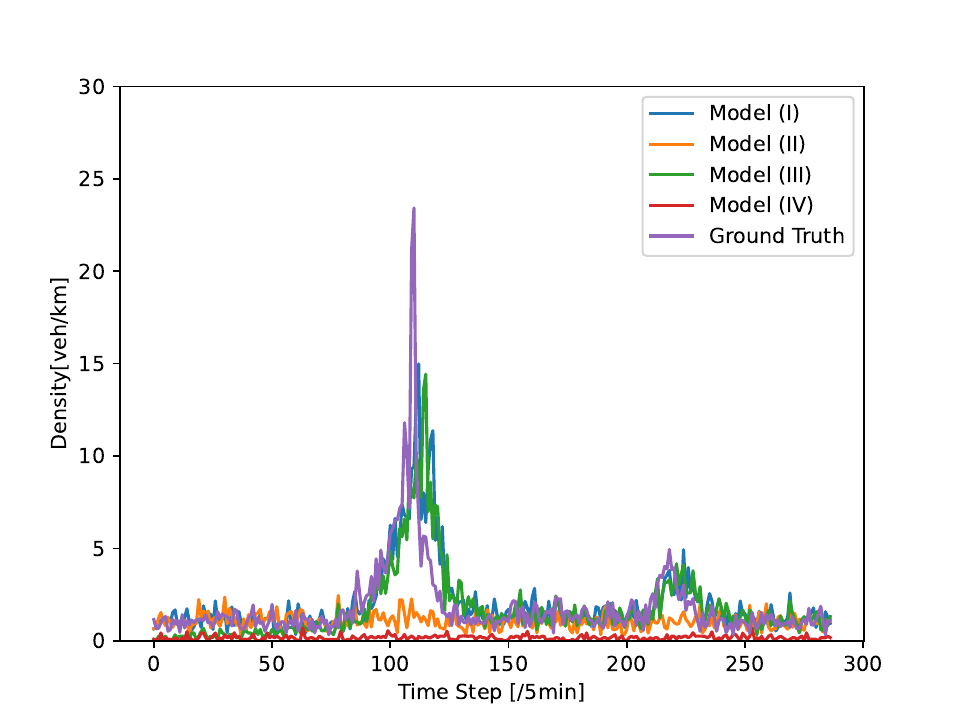}
    \caption{Location 1}
    \label{fig:Density_location_1}
\end{subfigure}
\begin{subfigure}[b]{0.5\columnwidth}
    \centering
    \includegraphics[width=\textwidth]{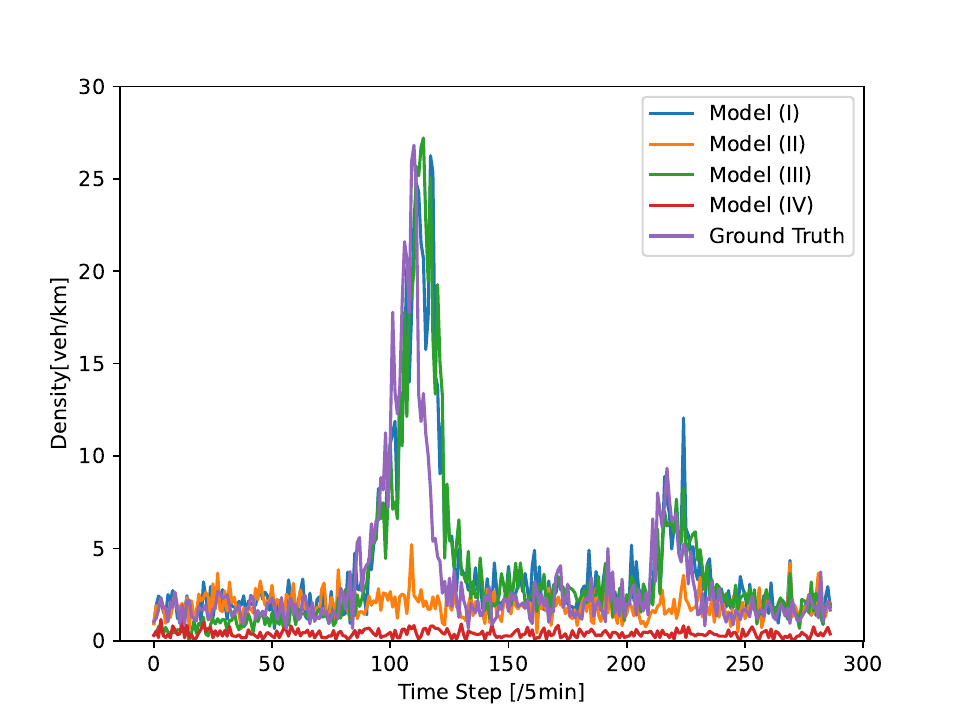}
    \caption{Location 2}
    \label{fig:Density_location_2}
\end{subfigure}%
\begin{subfigure}[b]{0.5\columnwidth}
    \centering
    \includegraphics[width=\textwidth]{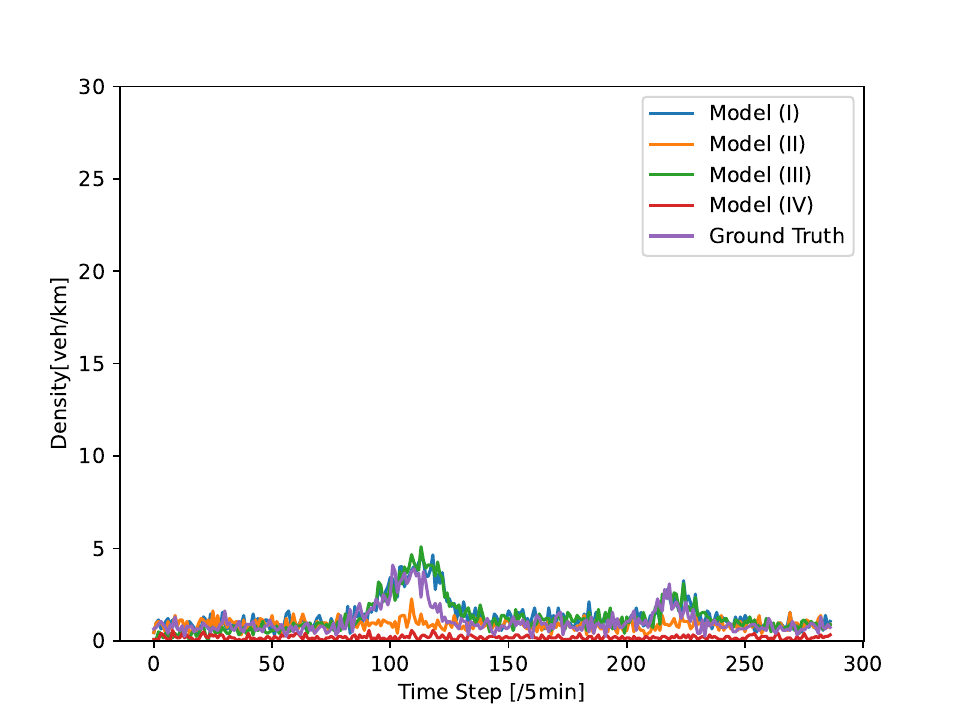}
    \caption{Location 3}
    \label{fig:Density_location_3}
\end{subfigure}
\caption{Time series of traffic demand at various locations}
\label{fig:Density_locations}
\end{figure}

We assess the performance of the proposed framework in an incident case where one lane on the main road’s link is closed. We compare the four models: 
\begin{itemize}
    \item Model (I): Uses parameters learnt by both one typical day simulation and pre-training, and updates the parameters in real-time. 
    \item Model (II): Uses parameters learnt by both one typical day simulation and pre-training,  but {\it does not} update the parameters in real-time during this incident case.
    \item Model (III): Uses parameters learnt only by pre-training and updates the parameters in real time.
    \item Model (IV): Uses parameters learnt only by pre-training,  but {\it does not} update the parameters in real-time during this incident case. 
  
\end{itemize}

\begin{figure}[p]
\centering
\begin{subfigure}[b]{0.5\columnwidth}
    \centering
    \includegraphics[width=\textwidth]{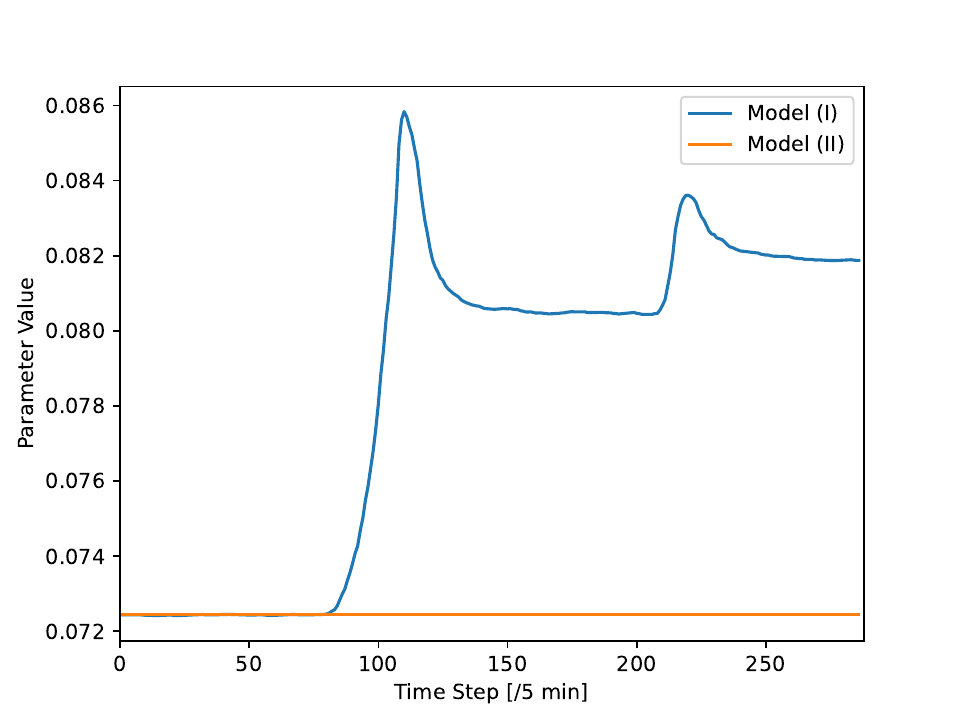}
    \caption{Parameter I}
    \label{fig:parameter0}
\end{subfigure}%
\begin{subfigure}[b]{0.5\columnwidth}
    \centering
    \includegraphics[width=\textwidth]{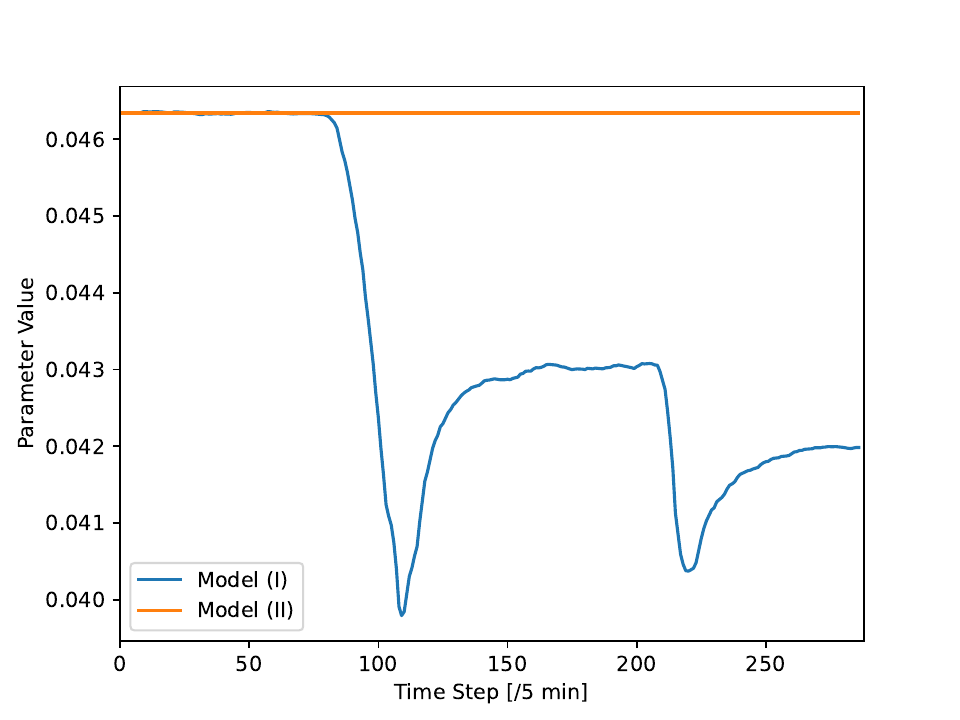}
    \caption{Parameter II}
    \label{fig:parameter1}
\end{subfigure}
\begin{subfigure}[b]{0.5\columnwidth}
    \centering
    \includegraphics[width=\textwidth]{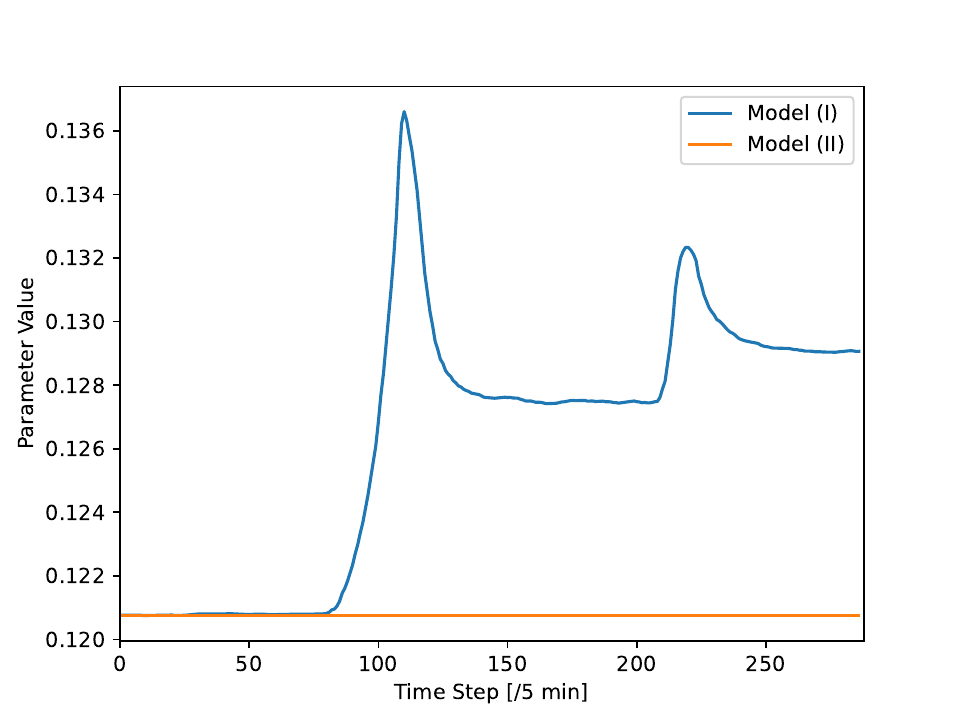}
    \caption{Parameter III}
    \label{fig:parameter2}
\end{subfigure}%
\begin{subfigure}[b]{0.5\columnwidth}
    \centering
    \includegraphics[width=\textwidth]{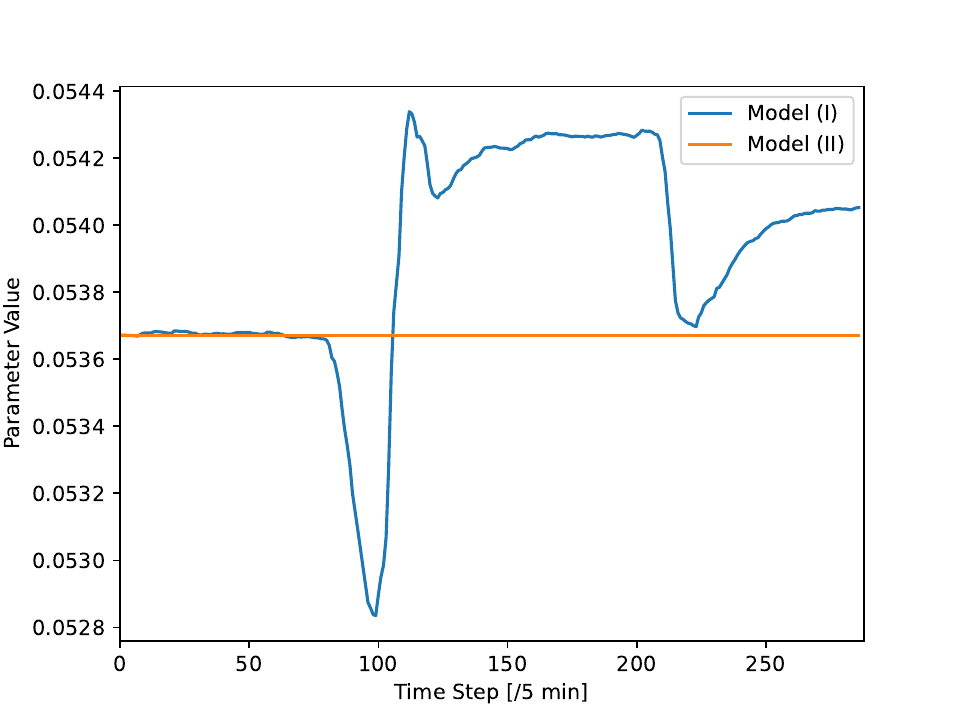}
    \caption{Parameter IV}
    \label{fig:parameter3}
\end{subfigure}
\begin{subfigure}[b]{0.5\columnwidth}
    \centering
    \includegraphics[width=\textwidth]{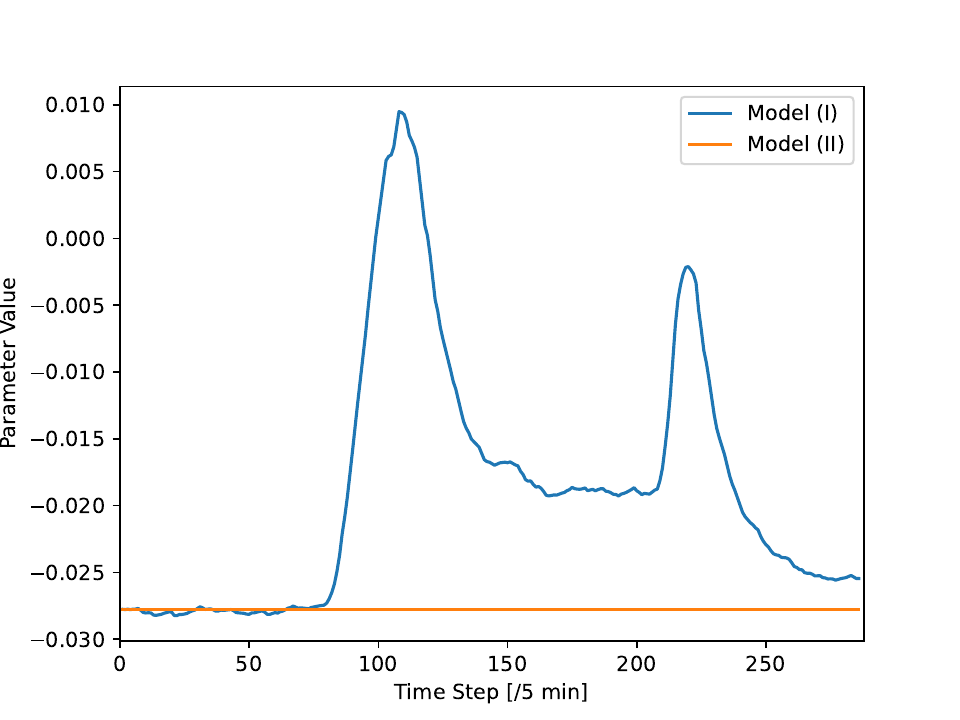}
    \caption{Parameter V}
    \label{fig:parameter4}
\end{subfigure}%
\begin{subfigure}[b]{0.5\columnwidth}
    \centering
    \includegraphics[width=\textwidth]{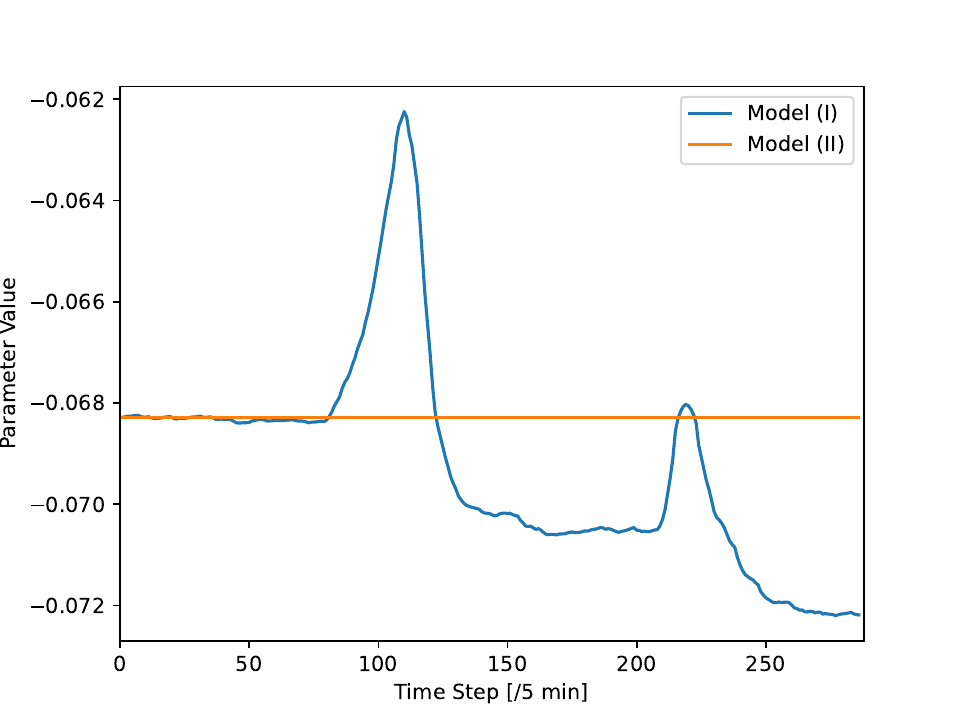}
    \caption{Parameter VI}
    \label{fig:parameter5}
\end{subfigure}
\caption{Time series of various parameters}
\label{fig:parameters}
\end{figure}

\begin{figure}[t]
\centering
\begin{subfigure}[b]{0.5\columnwidth}
    \centering
    \includegraphics[width=\textwidth]{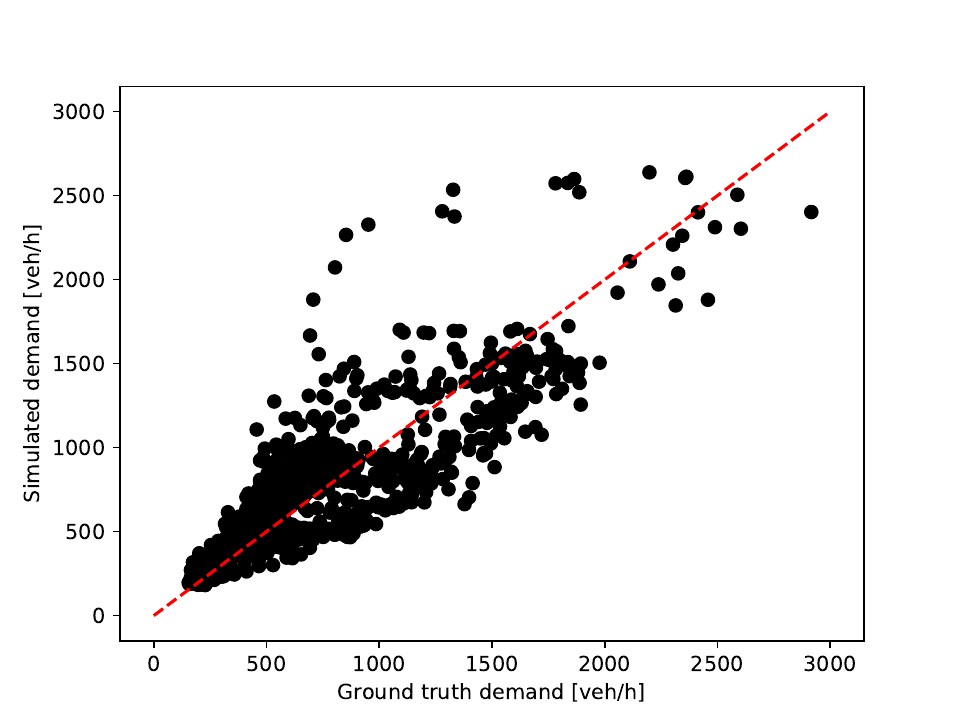}
    \caption{Model I}
    \label{fig:Demand_comparison_w_RT_from_learned_p}
\end{subfigure}%
\begin{subfigure}[b]{0.5\columnwidth}
    \centering
    \includegraphics[width=\textwidth]{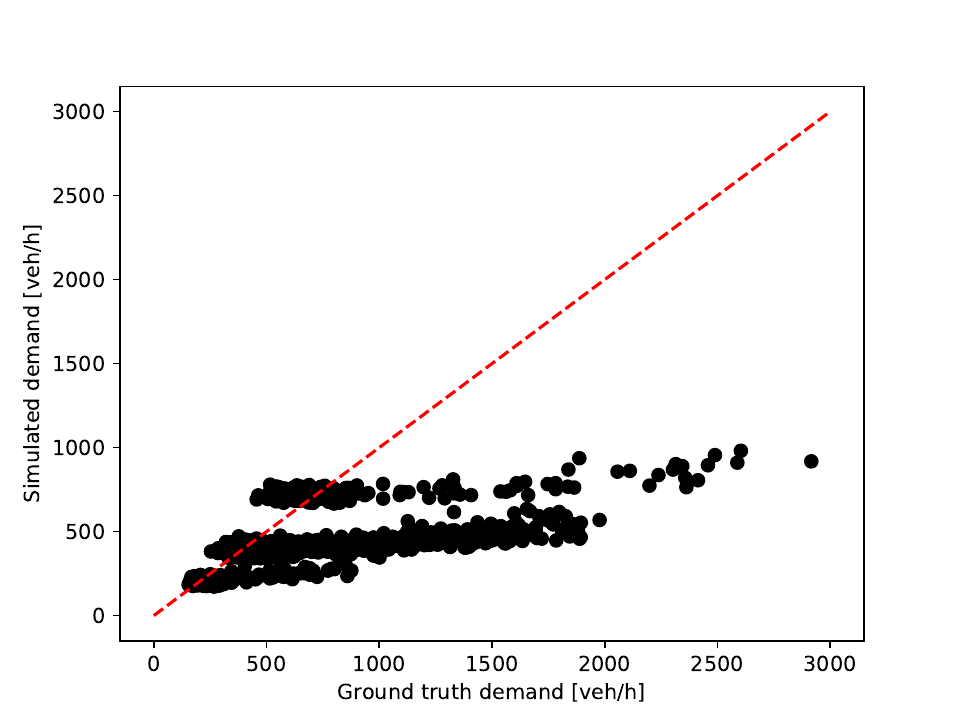}
    \caption{Model II}
    \label{fig:Demand_comparison_wo_RT_from_learned_p}
\end{subfigure}
\begin{subfigure}[b]{0.5\columnwidth}
    \centering
    \includegraphics[width=\textwidth]{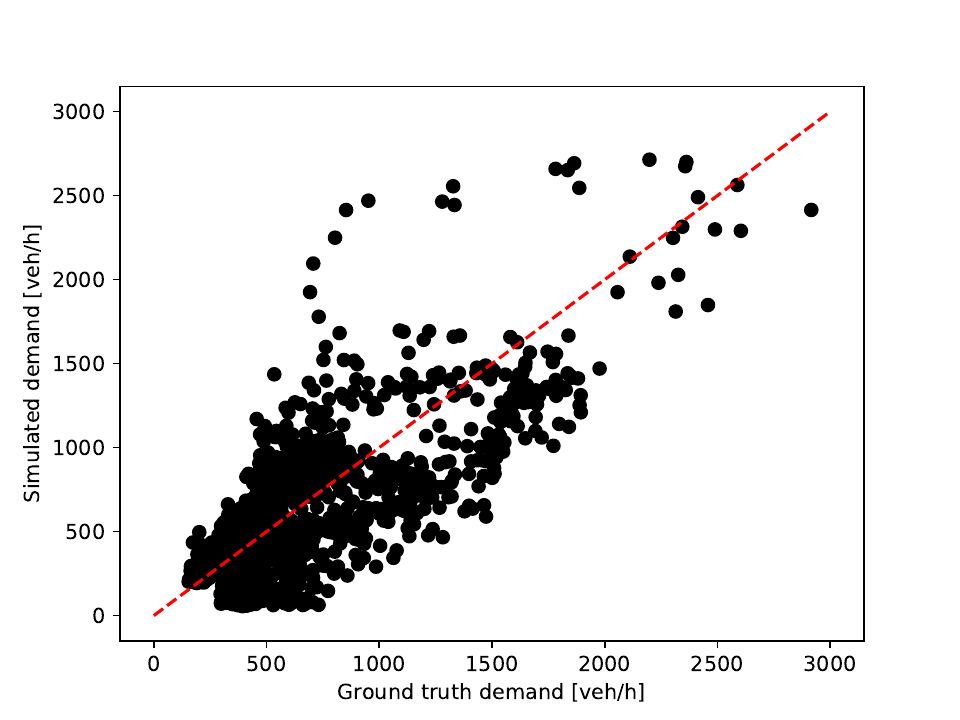}
    \caption{Model III}
    \label{fig:Demand_comparison_w_RT}
\end{subfigure}%
\begin{subfigure}[b]{0.5\columnwidth}
    \centering
    \includegraphics[width=\textwidth]{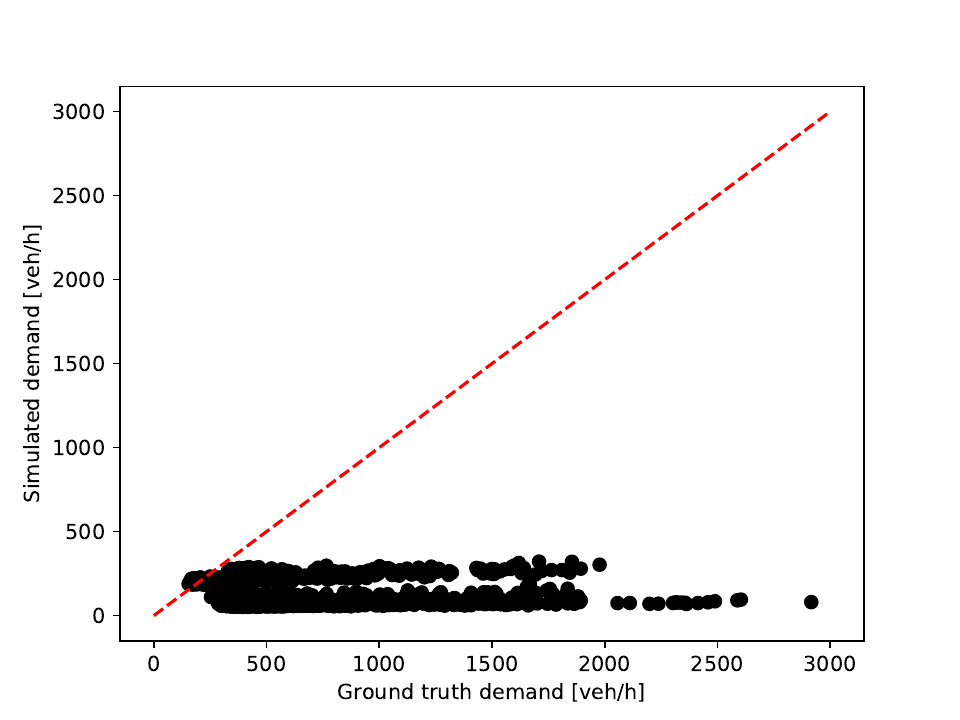}
    \caption{Model IV}
    \label{fig:Demand_comparison__wo_RT}
\end{subfigure}
\caption{Comparison between predicted and ground truth demand for different models}
\label{fig:Demand_comparison}
\end{figure}

\subsubsection{Results}
Fig.~\ref{fig:loss_toy} displaying the loss function over time shows that Model (I) outperforms the other models, particularly during the morning peak period. Model (IV) performs well during off-peak periods; however, its predictions deteriorate during the morning and evening peak periods. This is because of the static nature of the employed metamodel. Accurate predictions by the metamodel are feasible under undersaturated conditions, where traffic conditions between time steps are weakly correlated. In contrast, performance under oversaturated conditions deteriorates due to the influence of preceding traffic states on subsequent ones. Since the DNN parameters of Model (II) are updated using the traffic data of one typical day simulation in addition to the pre-training dataset, the performance of Model (II) is better than that of Model (IV). However, as traffic patterns of the incident case are different from those of the typical days, the performance especially during the peak periods deteriorates without the real-time update of the DNN parameters. This indicates the importance of the real-time update of the DNN parameters. Model (II) has slightly better performance than Model (IV), which could be because Model (II) has previous knowledge of oversaturated conditions of a typical day simulation.

The time series of traffic density at each measurement point in Figs.~\ref{fig:Density_location_0} -- \ref{fig:Density_location_3} also shows that Model (II) and (IV) performs well during the off-peak periods, but cannot capture the traffic patterns during the peak periods. We also observe that Models (I) and (III) have two different types of gaps from the ground-truth traffic density. One is that traffic conditions are underestimated during the loading periods of the peaks, and the other is that traffic conditions are overestimated during unloading periods. Since traffic congestion becomes more severe than usual when one of the lanes is closed, the models underestimate the traffic conditions at the beginning. 

Table~\ref{tab:mse_models} shows that Model (I) outperforms the other models at all locations in terms of MSE. The MSE of Models (II) and (IV) is large particularly at location 2, where the most severe congestion occurs during peak periods. As discussed above, models trained only with the static metamodel cannot cope with oversaturated conditions.

As seen in Figs.~\ref{fig:parameter0} -- \ref{fig:parameter5} showing the time series of DNN parameters in Model (I), the DNN model parameters are updated to fill the gaps between the simulated and ground-truth traffic density. As a result, the gap is mitigated at the peak, but traffic conditions are overestimated during the unloading periods. Again, the DNN model parameters are updated to fill the gaps as shown in Figs.~\ref{fig:parameter0} -- \ref{fig:parameter5}.

Finally, we compare the predicted traffic demand with the ground-truth traffic demand, as depicted in 
 Figs.~\ref{fig:Demand_comparison_w_RT_from_learned_p} -- \ref{fig:Demand_comparison__wo_RT}.  Model (I) and (III) can predict not only low traffic demand but also high traffic demand, while Model (II) and (IV) can predict only low traffic demand. As discussed above, Model (II) and (IV) fail to predict accurate traffic demand during peak periods. Furthermore, we can see that the predicted traffic demand from Model (I) is closer to the ground-truth traffic demand than that from Model (III), which could be because of the previous knowledge of the oversaturated traffic conditions from the typical day simulation. 

 Overall, these observations confirms the effectiveness of the proposed real-time simulation framework (i.e., Model (I)), even in incident case where traffic patterns change significantly.

\subsection{Tokyo network}
\subsubsection{Simulation set-up}
\begin{figure}[t]
\centering
\includegraphics[width=\textwidth]{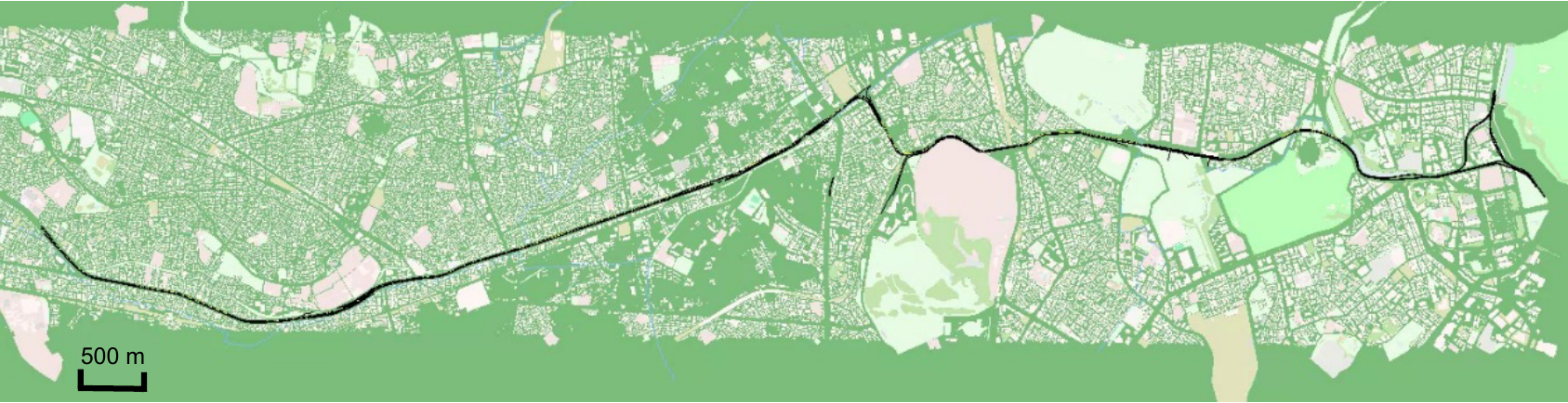}
\caption{Tokyo expressway network. Black lines are the network for consideration}
\label{fig:tokyo_network}
\end{figure}

We next demonstrate the performance of the proposed framework in a large-scale network with real data, specifically the Tokyo expressway corridor network under recurrent and non-recurrent cases.  The network for consideration is depicted in Fig.~\ref{fig:tokyo_network} and is imported from \cite{OP}. The two-lane road length is 13.3 km with 6 loop detectors and there are 27 OD pairs in total.  The loop detector data contains flow and speed collected every 5 min.  The data spans from July 1, 2018 to September 14, 2018, and July 19, 2021.  The 2018 July and August data is used for pre-training, while the September data is used for recurrent cases. Additionally, data from July 19, 2021, when travel demand management for the Tokyo 2020 Olympic and Paralympic Games started, is used to represent non-recurrent cases. \cite{dantsuji2024} found that a 1000 JPY additional toll surcharges  decreased traffic demand of passenger vehicles by 25.0 $\%$ during the Games. Moreover, July 19, 2021, is the first day of the TDM implementation, presenting a clear example of non-recurrent traffic patterns.  

To accurately reflect the propagation of congestion, we introduce ``ghost vehicles" that travel at the same speed as those recorded by the loop detectors on the upstream links of the junctions at the network’s boundaries. Additionally, we adjust the time gap values in the car-following model at the sag differently from those on normal roads to account for its impacts.

We use the same metamodel as in Eq. (\ref{eq:density_estimation}). For the inputs of the DNN, we utilize the timestep, the day of the week, and the dummy variable of weekends/holidays as one-hot encoding time-related variables, along with both flow and speed as traffic measurement variables. The DNN architecture is Long Short-Term Memory (LSTM). The LSTM architecture consists of an LSTM layer followed by a fully connected (dense) output layer. The LSTM layer has a hidden dimension of 128 units and 2 layers.  The output of the LSTM layer is fed into a fully connected layer with a linear activation function

\subsubsection{Results: the recurrent case}
\begin{figure}[p]
\begin{minipage}{\columnwidth}
\centering
\includegraphics[width=\textwidth]{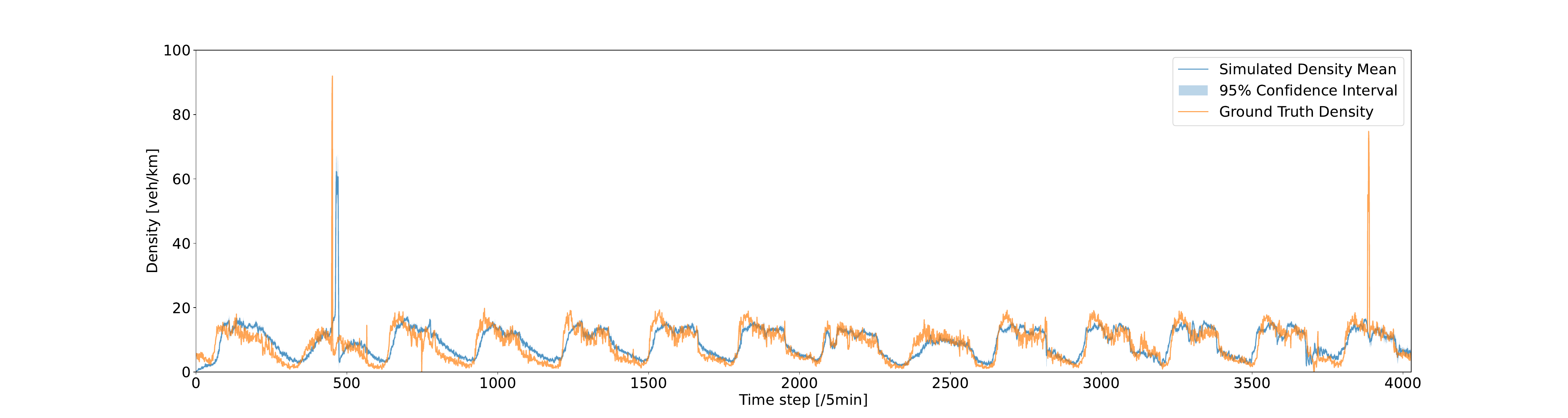}
\caption{Time series of traffic density at a uncongested location }
\label{fig:Location_1_recurrent}
\end{minipage}
\begin{minipage}{\columnwidth}
\centering
\includegraphics[width=\textwidth]{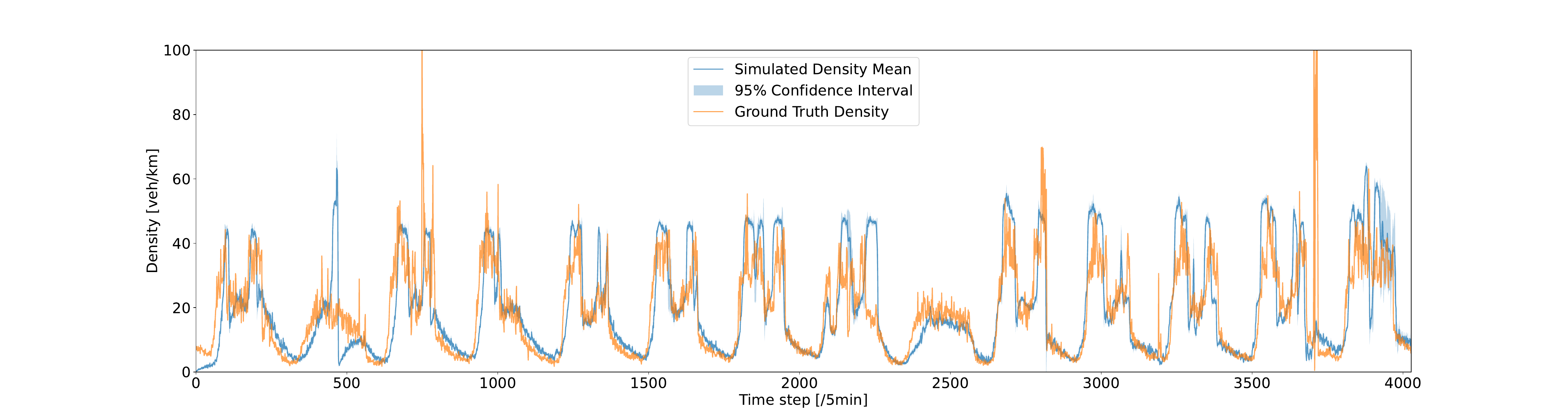}
\caption{Time series of traffic density at a congested location}
\label{fig:Location_2_recurrent}
\end{minipage}
\begin{minipage}{\columnwidth}
\centering
\includegraphics[width=\textwidth]{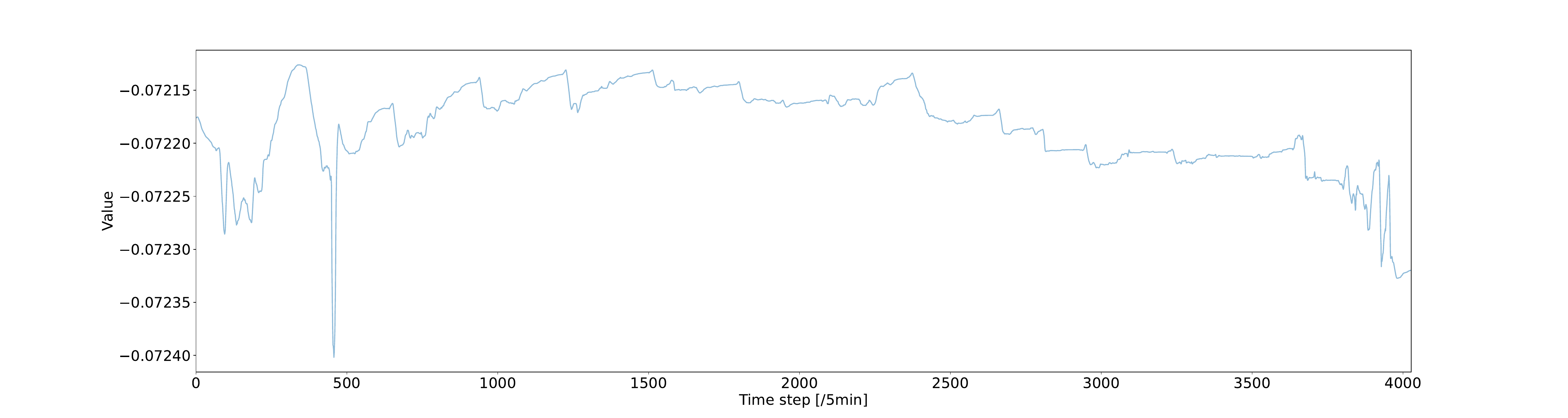}
\caption{Time series of LSTM parameter I}
\label{fig:Parameter_0_recurrent}
\end{minipage}
\begin{minipage}{\columnwidth}
\centering
\includegraphics[width=\textwidth]{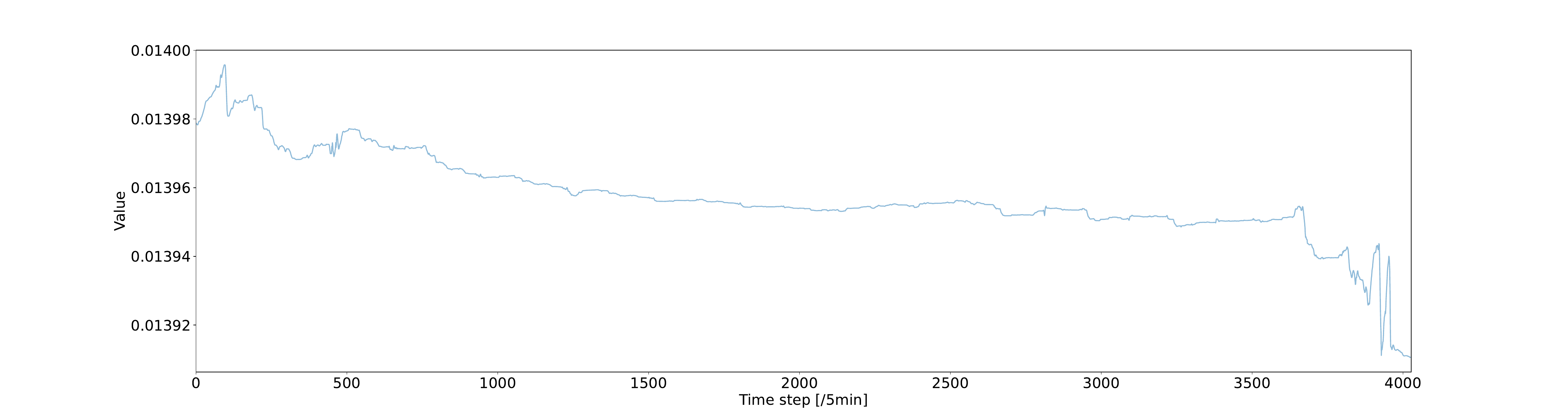}
\caption{Time series of LSTM parameter II}
\label{fig:Parameter_1_recurrent}
\end{minipage}
\end{figure}

We first investigate the performance of the proposed framework under recurrent traffic conditions. As depicted in Figs~\ref{fig:Location_1_recurrent} and \ref{fig:Location_2_recurrent}, the time series of traffic density of traffic density at both congested and uncongested locations for two weeks demonstrate the good performance of the proposed method. The simulation accurately captures the recurrent traffic patterns over time, particularly at the uncongested location at the beginning, which demonstrates the efficacy of the offline pre-training process. However, due to the static nature of the metamodel, gaps between simulated and actual traffic density at the congested location are observed during peak periods over some initial days, as depicted in Fig~\ref{fig:Location_2_recurrent}. These gaps diminish as the NN updates its parameters, learning from congestion data over several days. Note that we simply set up the lower time gap value in the car-following model at the sag, however, the car-following behavior at the sag is more complicated \cite[e.g.,][]{wada2020continuum}. The results are expected to be improved if we incorporate such models.

The update process of some DNN parameters is illustrated in Figs. \ref{fig:Parameter_0_recurrent} and \ref{fig:Parameter_1_recurrent}. During the first day (Saturday), small parameter changes are observed as the simulation captures the traffic patterns. However, the parameters change significantly before time step 500 on day 2. This may be because the DNN expects two peaks during the day, but there is only one small peak on Sunday, as depicted in Fig. \ref{fig:Location_2_recurrent}. The DNN then updates its parameters to learn this pattern and can reflect it on day 9 (around time step 2500).

\begin{figure}[t]
\centering
\includegraphics[width=\textwidth]{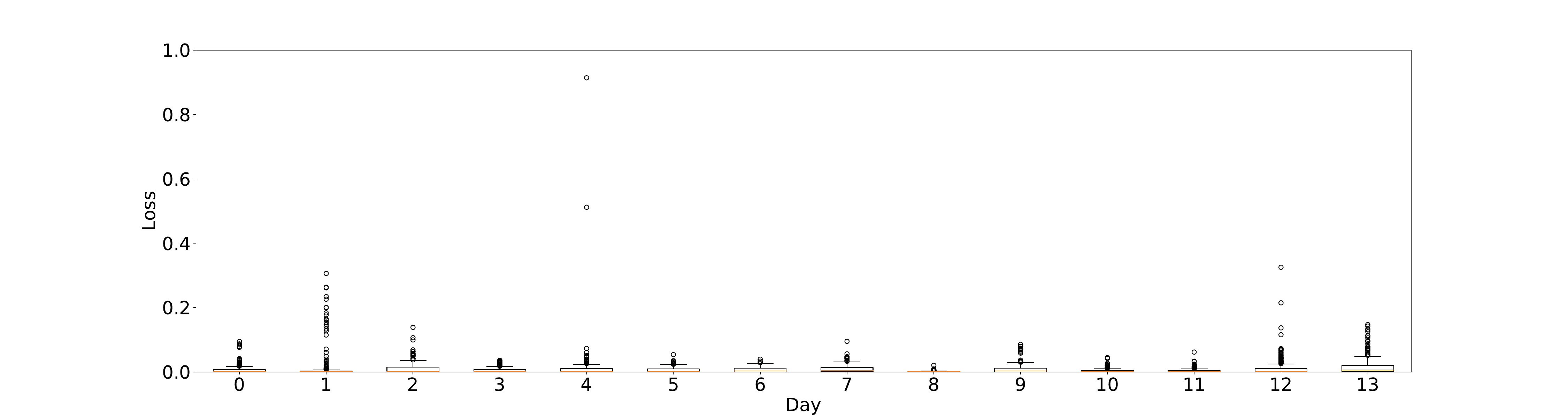}
\caption{Box plot of loss function regarding traffic measurements ($\mathcal{L}_f$) over days. Note that one sample from day 2 is excluded due to too high value}
\label{fig:Loss_tokyo}
\end{figure}

The box plot shown in Fig. \ref{fig:Loss_tokyo} summarizes the loss values for each day across different time steps. While there are certain periods where the loss values spike, indicating larger discrepancies between the simulation and the ground truth, it is evident that for the majority of the time, the simulation closely matches the actual data, resulting in low loss values. Note that the high loss values are attributed to errors in loop detector measurements. 

\begin{figure}[p]
\centering
\begin{subfigure}[b]{0.5\columnwidth}
    \centering
    \includegraphics[width=\textwidth]{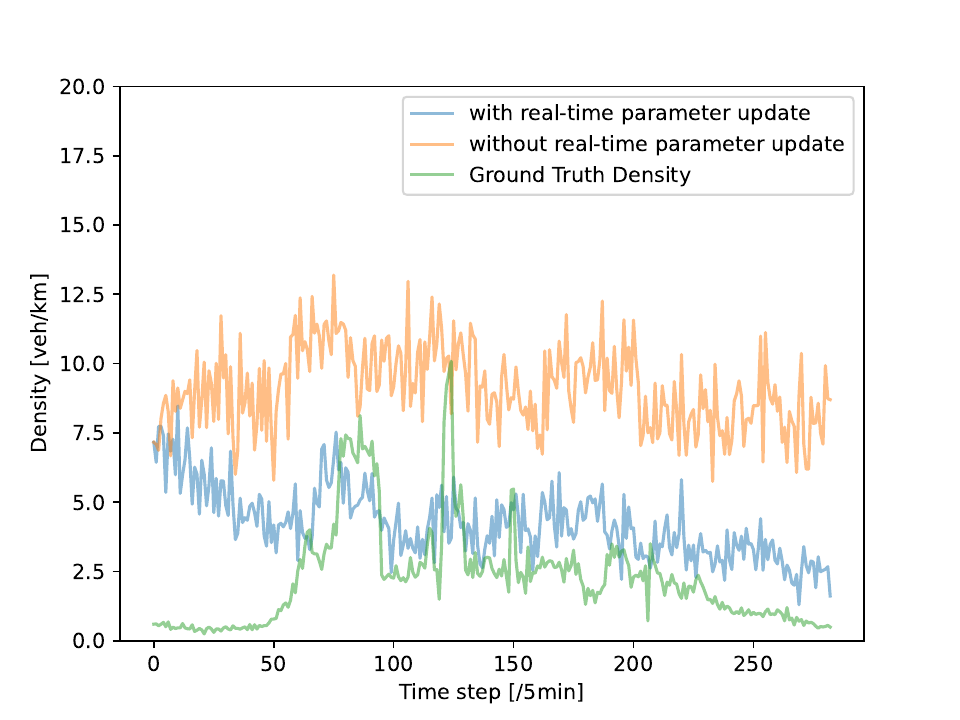}
    \caption{Location 1}
    \label{fig:Location_0_Oly}
\end{subfigure}%
\begin{subfigure}[b]{0.5\columnwidth}
    \centering
    \includegraphics[width=\textwidth]{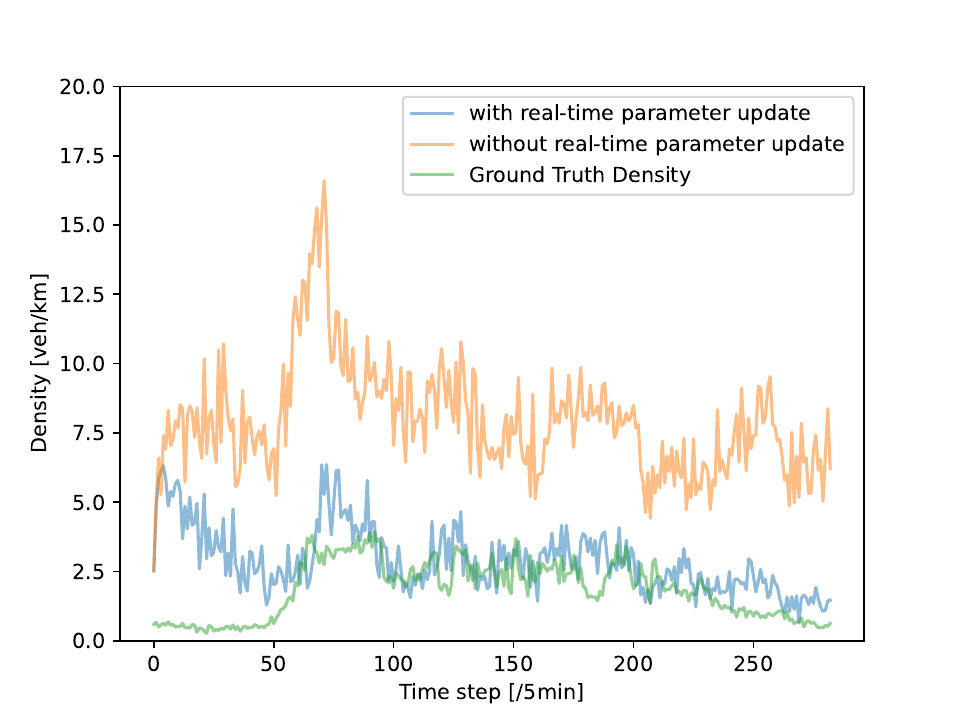}
    \caption{Location 2}
    \label{fig:Location_1_Oly}
\end{subfigure}
\begin{subfigure}[b]{0.5\columnwidth}
    \centering
    \includegraphics[width=\textwidth]{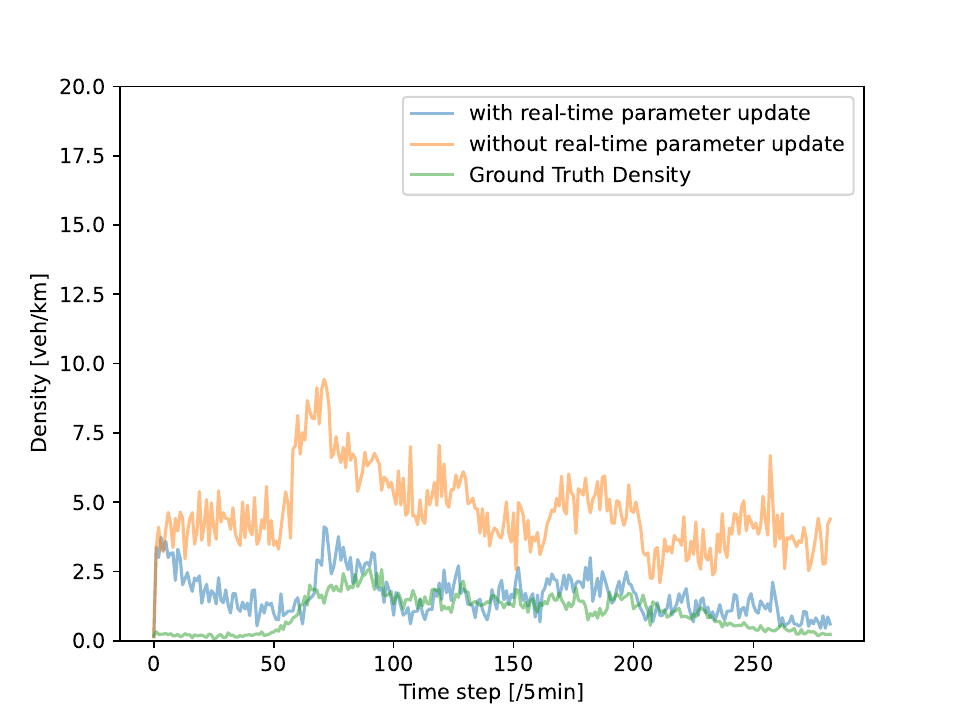}
    \caption{Location 3}
    \label{fig:Location_2_Oly}
\end{subfigure}%
\begin{subfigure}[b]{0.5\columnwidth}
    \centering
    \includegraphics[width=\textwidth]{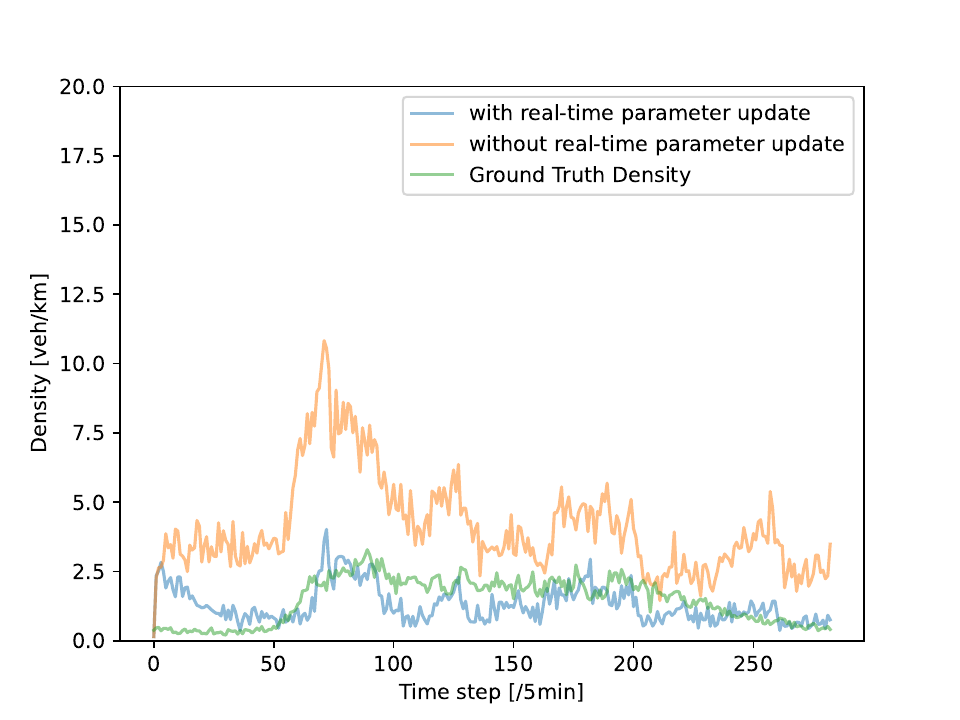}
    \caption{Location 4}
    \label{fig:Location_3_Oly}
\end{subfigure}
\begin{subfigure}[b]{0.5\columnwidth}
    \centering
    \includegraphics[width=\textwidth]{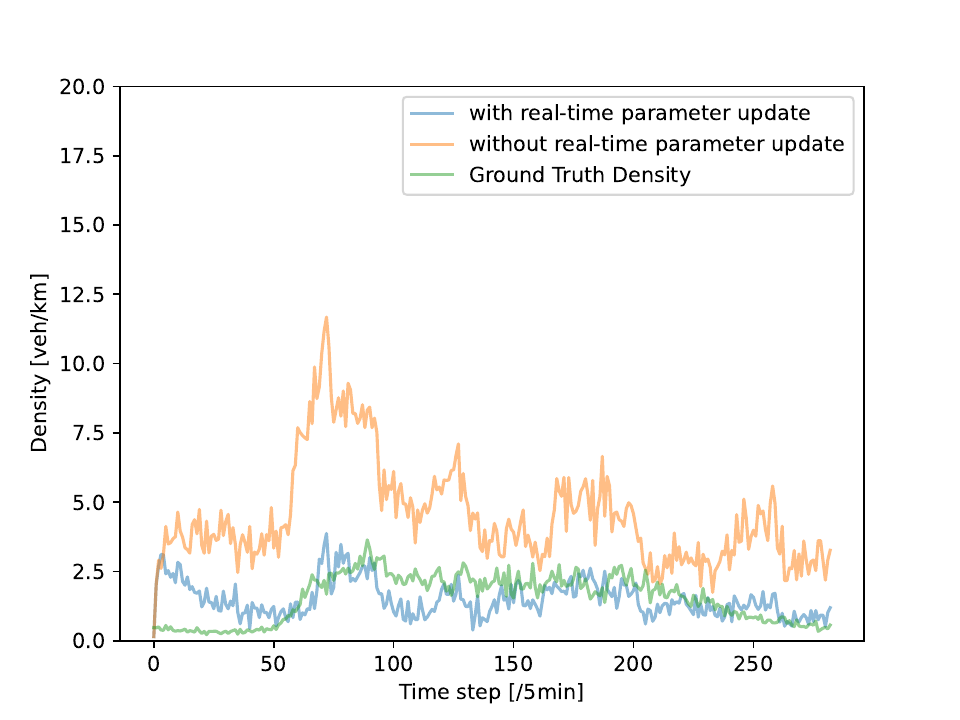}
    \caption{Location 5}
    \label{fig:Location_4_Oly}
\end{subfigure}%
\begin{subfigure}[b]{0.5\columnwidth}
    \centering
    \includegraphics[width=\textwidth]{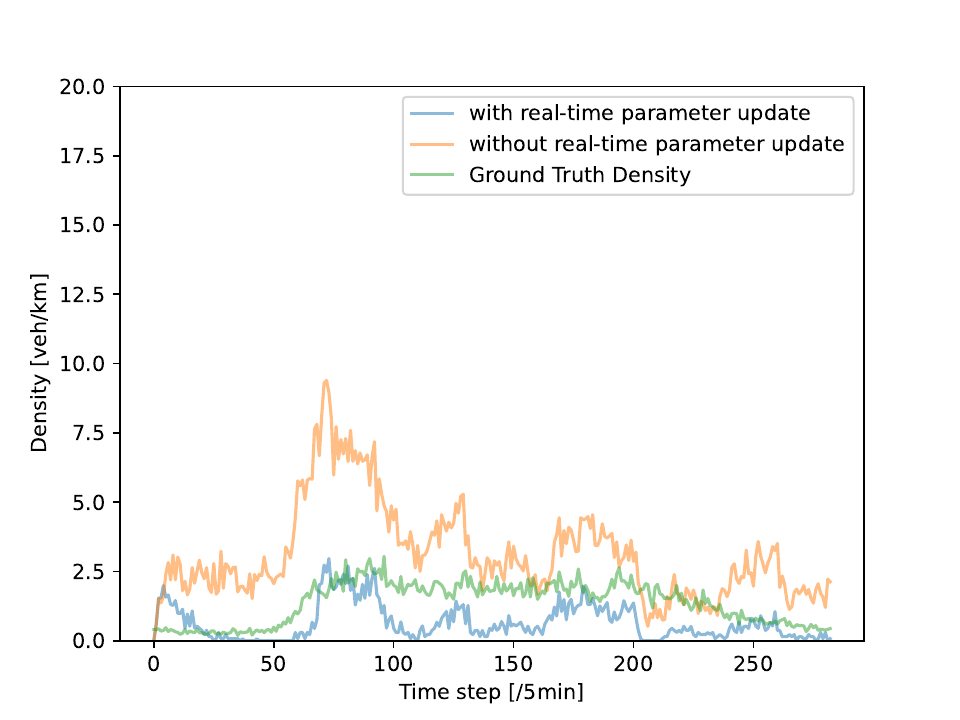}
    \caption{Location 6}
    \label{fig:Location_5_Oly}
\end{subfigure}
\caption{Time series of traffic density at various locations}
\label{fig:locations_oly}
\end{figure}

\subsubsection{Results: the non-recurrent case}
Using the DNN parameters learned from the recurrent case, we next evaluate the framework's performance under non-recurrent traffic conditions, such as those experienced during the Tokyo 2020 Olympic and Paralympic Games. To demonstrate the efficacy of real-time parameter updates, we compare models with and without these updates.

Figs~\ref{fig:Location_0_Oly}--\ref{fig:Location_5_Oly} display the time series of traffic density at each location. The  blue, orange, and green lines correspond to the models with and without the real-time parameter updates, and the ground-truth, respectively.  Without real-time parameter updates, the model fails to accurately capture traffic demand under non-recurrent conditions, overestimating traffic based on recurrent data knowledge. Conversely, the model with real-time parameter updates effectively adapts to changes in traffic demand during the Games. Despite initial overestimations of the traffic conditions, continuous updates enable the model to gradually learn and adjust to new traffic patterns, highlighting the importance of real-time parameter adjustments.  As shown in Table.~\ref{tab:mse_comparison}, the MSE of the model with the real-time parameter update is much lower than that of the model without the real-time parameter update.

\begin{figure}[p]
\centering
\begin{subfigure}[b]{0.5\columnwidth}
    \centering
    \includegraphics[width=\textwidth]{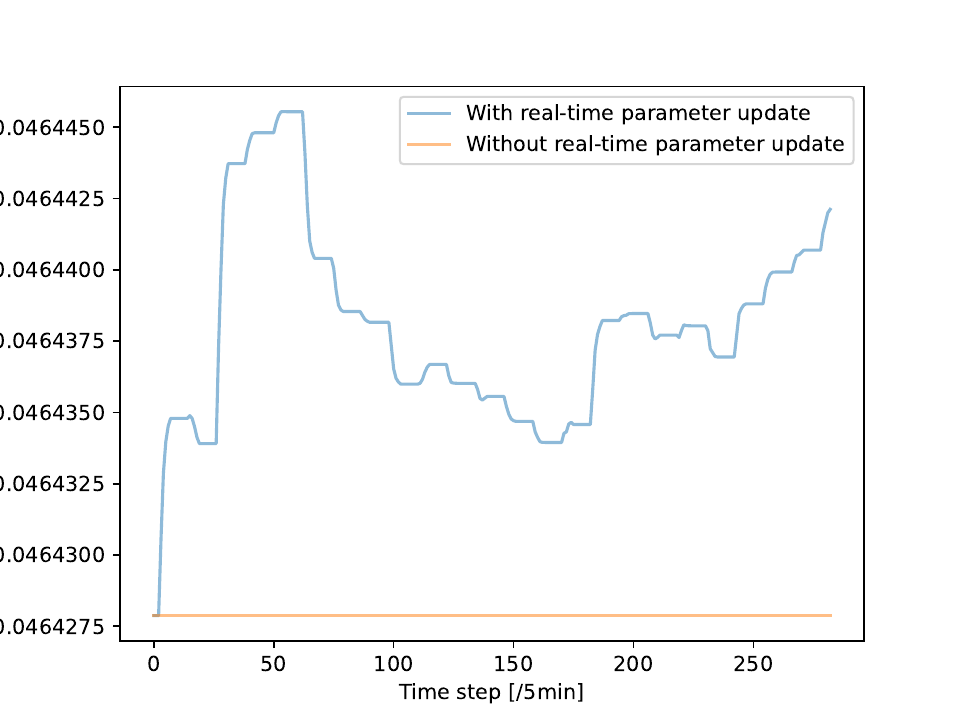}
    \caption{Parameter I}
    \label{fig:Parameter_45_Oly}
\end{subfigure}%
\begin{subfigure}[b]{0.5\columnwidth}
    \centering
    \includegraphics[width=\textwidth]{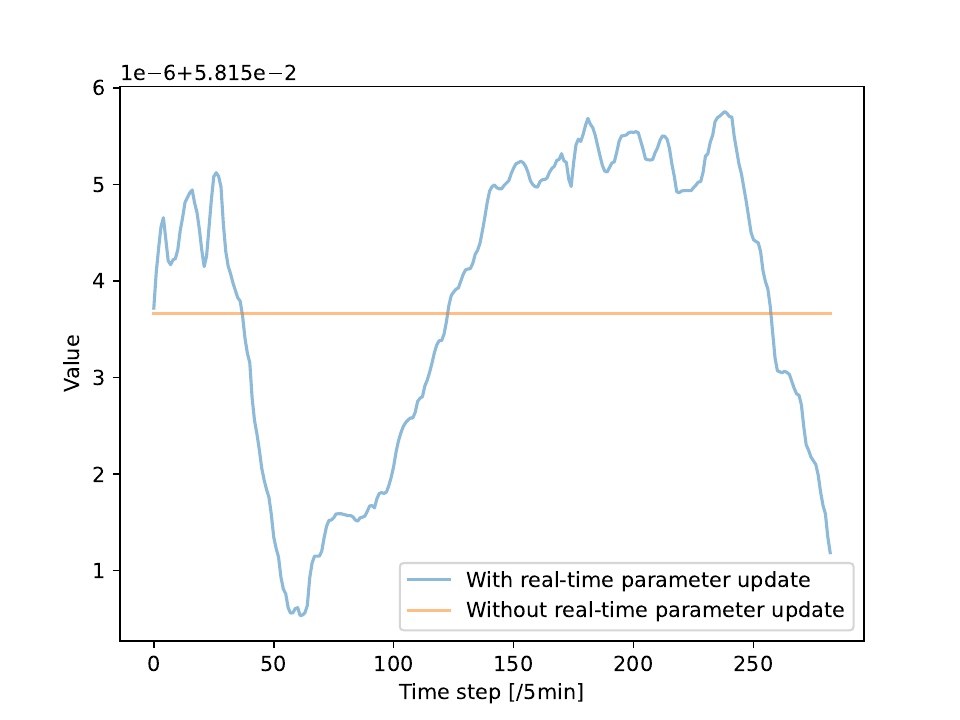}
    \caption{Parameter II}
    \label{fig:Parameter_97_Oly}
\end{subfigure}
\begin{subfigure}[b]{0.5\columnwidth}
    \centering
    \includegraphics[width=\textwidth]{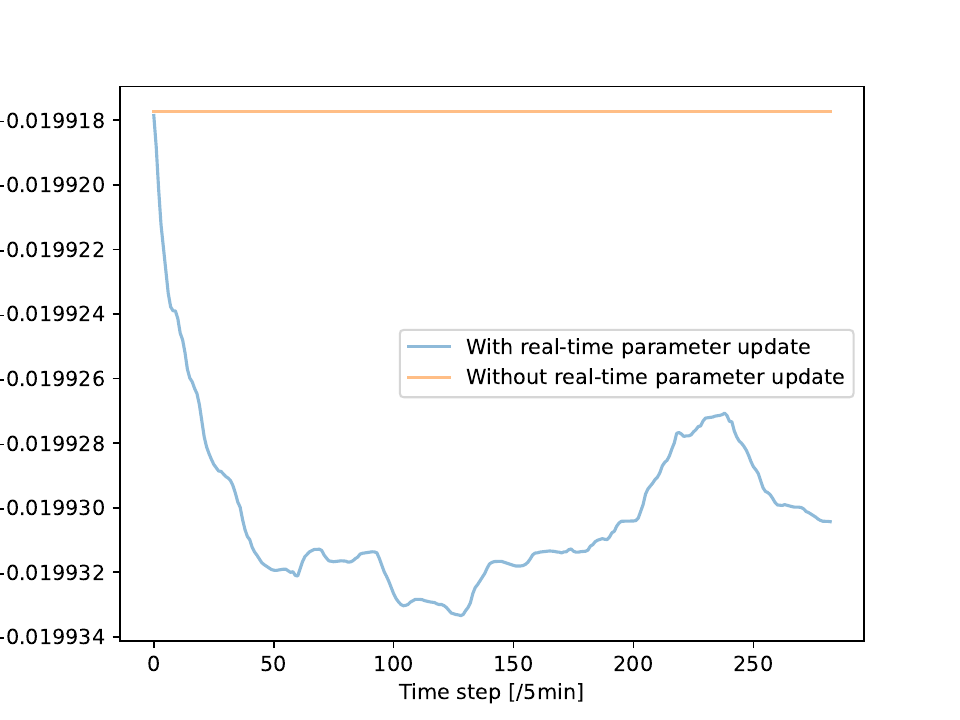}
    \caption{Parameter III}
    \label{fig:Parameter_101_Oly}
\end{subfigure}%
\begin{subfigure}[b]{0.5\columnwidth}
    \centering
    \includegraphics[width=\textwidth]{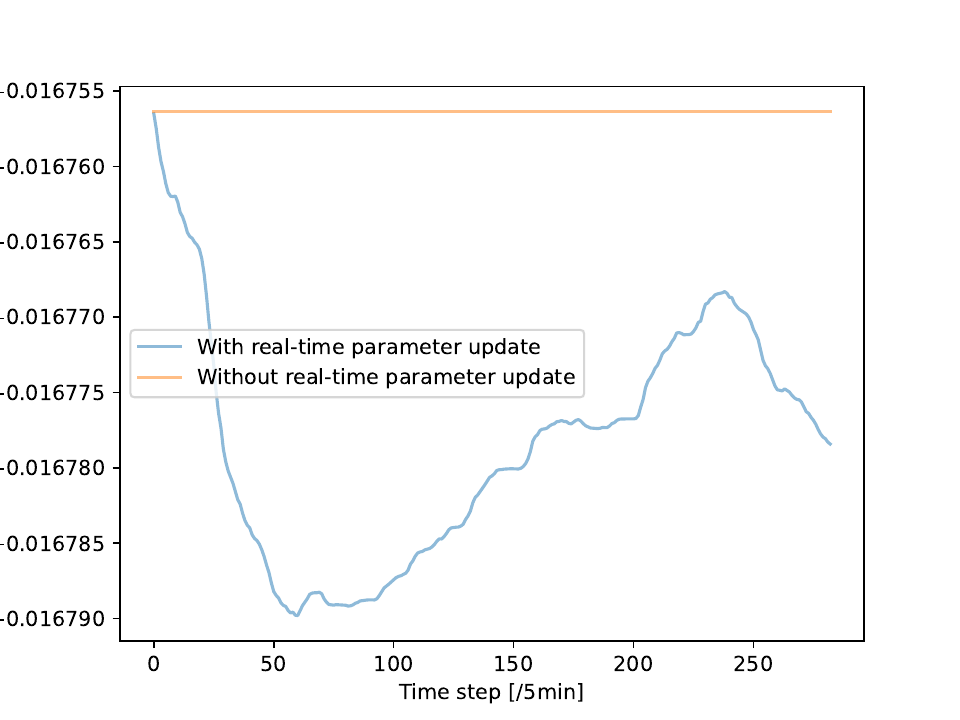}
    \caption{Parameter IV}
    \label{fig:Parameter_154_Oly}
\end{subfigure}
\begin{subfigure}[b]{0.5\columnwidth}
    \centering
    \includegraphics[width=\textwidth]{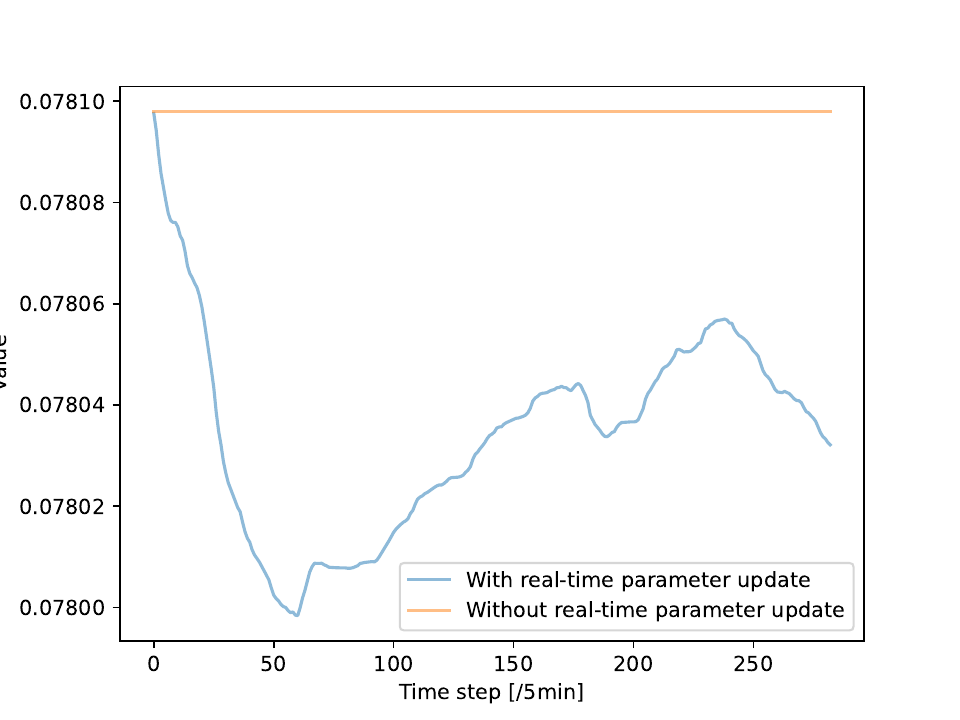}
    \caption{Parameter V}
    \label{fig:Parameter_157_Oly}
\end{subfigure}%
\begin{subfigure}[b]{0.5\columnwidth}
    \centering
    \includegraphics[width=\textwidth]{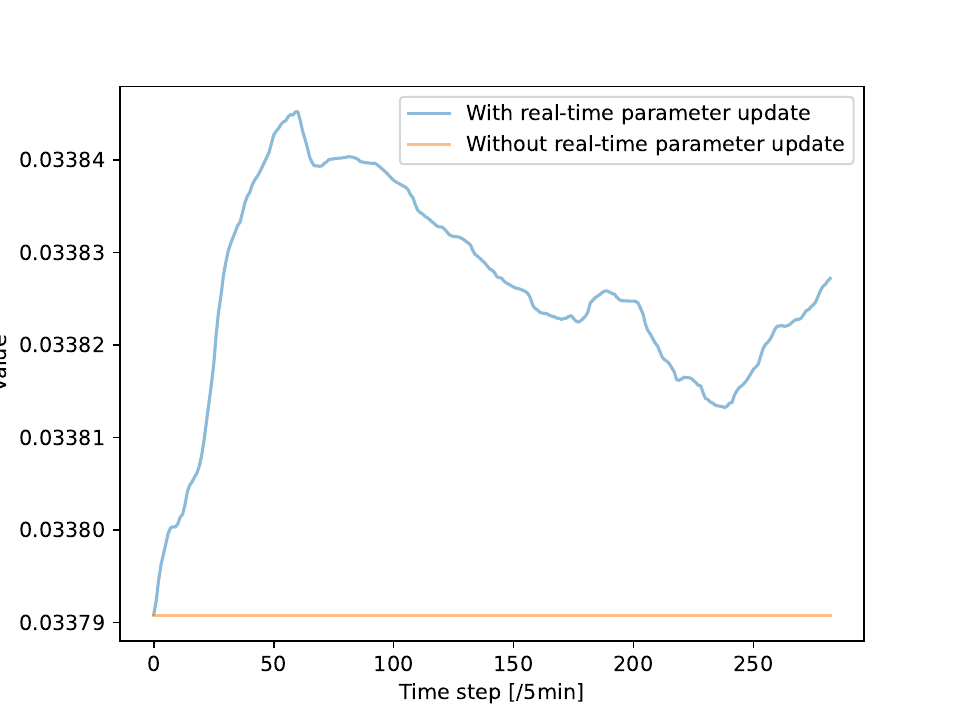}
    \caption{Parameter VI}
    \label{fig:Parameter_169_Oly}
\end{subfigure}
\caption{Time series of various parameters}
\label{fig:parameters_oly}
\end{figure}

The time series of selected parameter values (Figs~\ref{fig:Parameter_45_Oly}--\ref{fig:Parameter_169_Oly})  illustrate the continuous updating process, demonstrating  how the model fills the gaps between simulated and actual traffic conditions. We can see that the parameters adjust continuously to close these gaps.

\begin{table}[h]
    \centering
        \caption{Mean Squared Error (MSE) for different locations.}
    \label{tab:mse_comparison}
    \begin{tabular}{ccc}
        \textbf{Location} & \textbf{MSE with update} & \textbf{MSE without update} \\
        \hline
        0 & 8.3 & 50.3 \\
        1 & 3.0 & 40.7 \\
        2 & 1.0 & 14.6 \\
        3 & 0.7 & 9.2 \\
        4 & 0.8 & 11.4 \\
        5 & 1.1 & 5.3 \\
        \hline
        && [(veh/km)$^2$]
    \end{tabular}

\end{table}

\subsection{Sensitivity analysis}
We conduct the sensitivity analysis with respect to the importance of a priori demand (i.e., $\delta$ in Eq.~(\ref{eq:obj_function})) since the predefined $\delta$ must be set carefully if the quality of the result highly depends on this factor.  We set $\delta=\{ 0, 0.001, 0.1, 1\}$ for the non-recurrent case in the Tokyo network. A $\delta=0$  indicates that a priori demand has no importance for OD calibration, while a  $\delta=1$ suggests that a priori demand is as important as traffic states.

\begin{figure}[p]
\centering
\begin{subfigure}[b]{0.5\columnwidth}
    \centering
    \includegraphics[width=\textwidth]{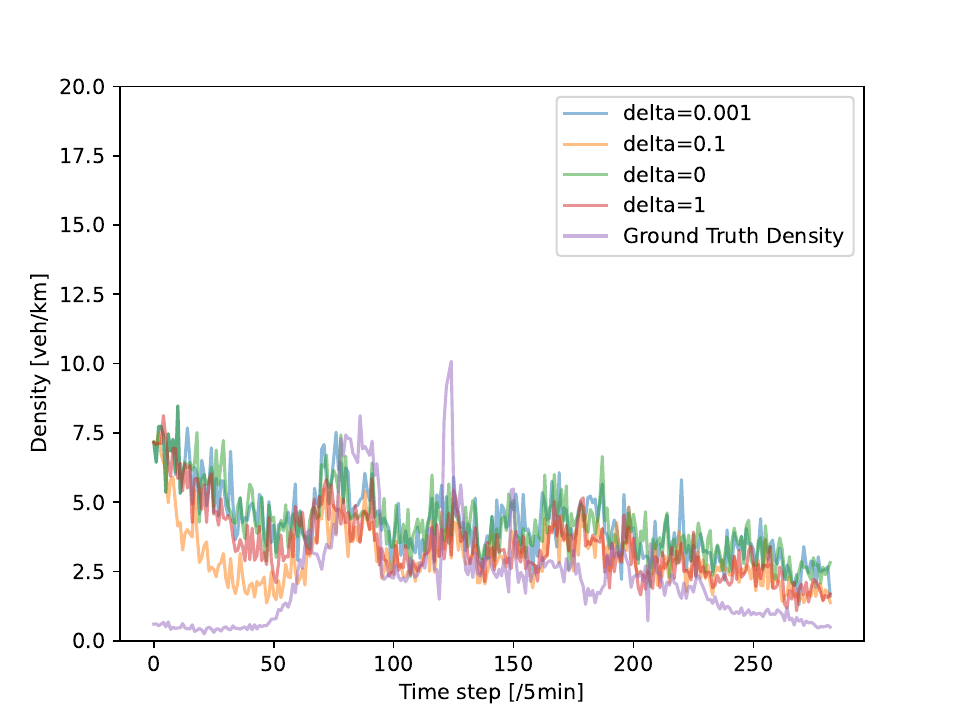}
    \caption{Location 1}
    \label{fig:Location_0_Oly_sensitivity}
\end{subfigure}%
\begin{subfigure}[b]{0.5\columnwidth}
    \centering
    \includegraphics[width=\textwidth]{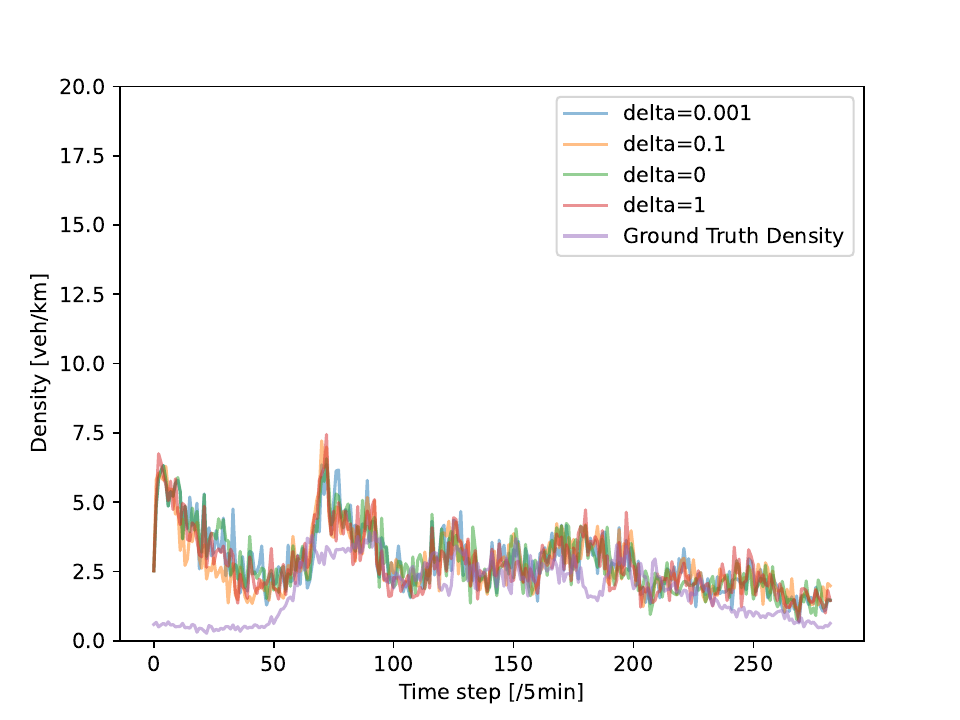}
    \caption{Location 2}
    \label{fig:Location_1_Oly_sensitivity}
\end{subfigure}
\begin{subfigure}[b]{0.5\columnwidth}
    \centering
    \includegraphics[width=\textwidth]{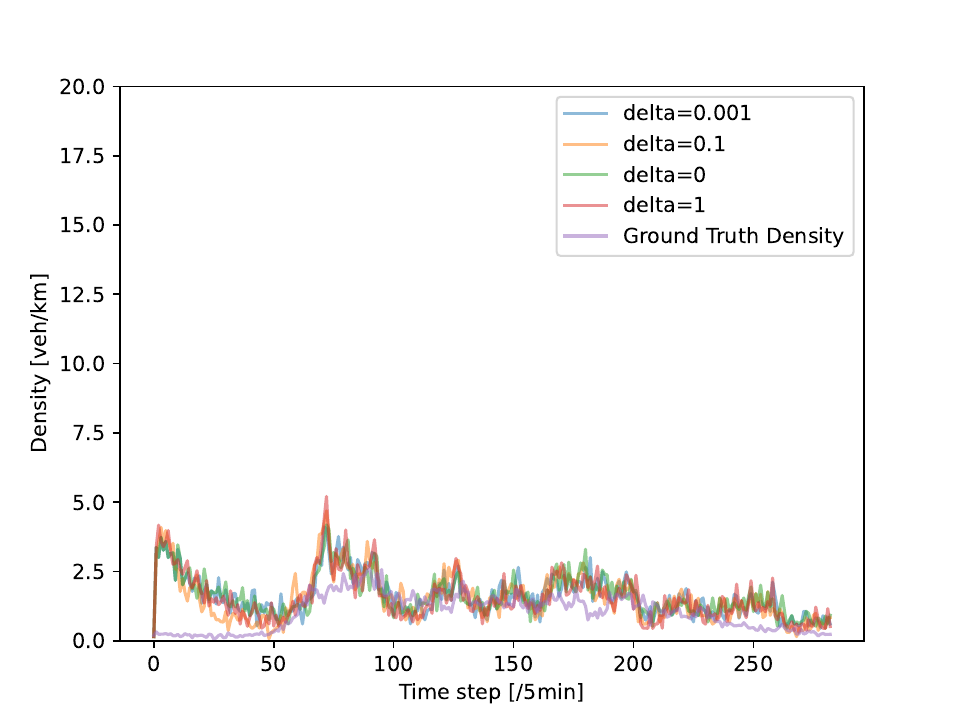}
    \caption{Location 3}
    \label{fig:Location_2_Oly_sensitivity}
\end{subfigure}%
\begin{subfigure}[b]{0.5\columnwidth}
    \centering
    \includegraphics[width=\textwidth]{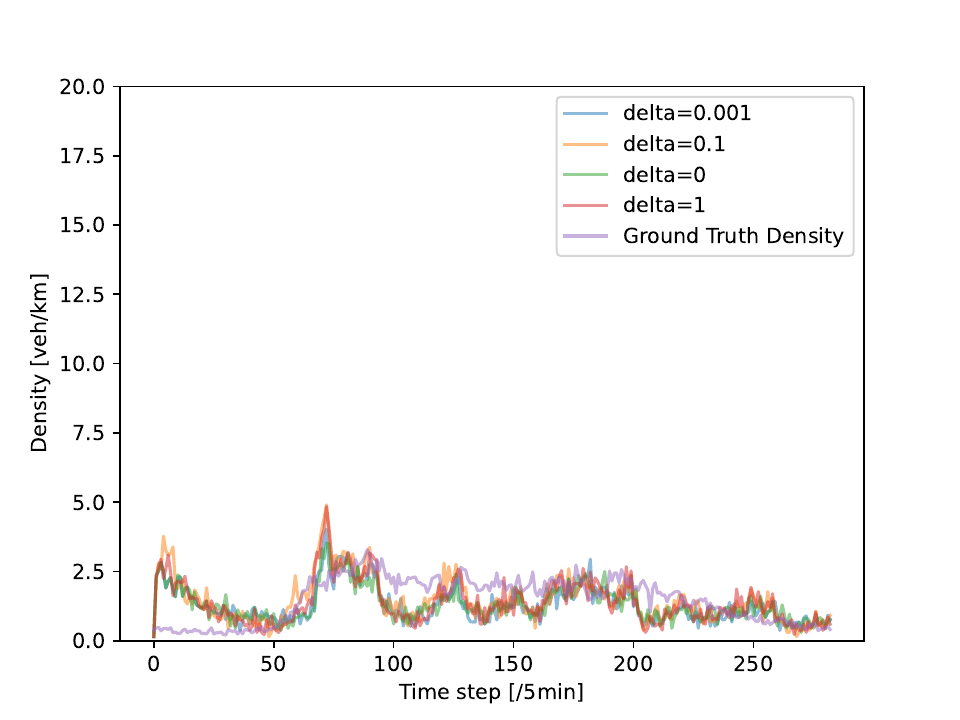}
    \caption{Location 4}
    \label{fig:Location_3_Oly_sensitivity}
\end{subfigure}
\begin{subfigure}[b]{0.5\columnwidth}
    \centering
    \includegraphics[width=\textwidth]{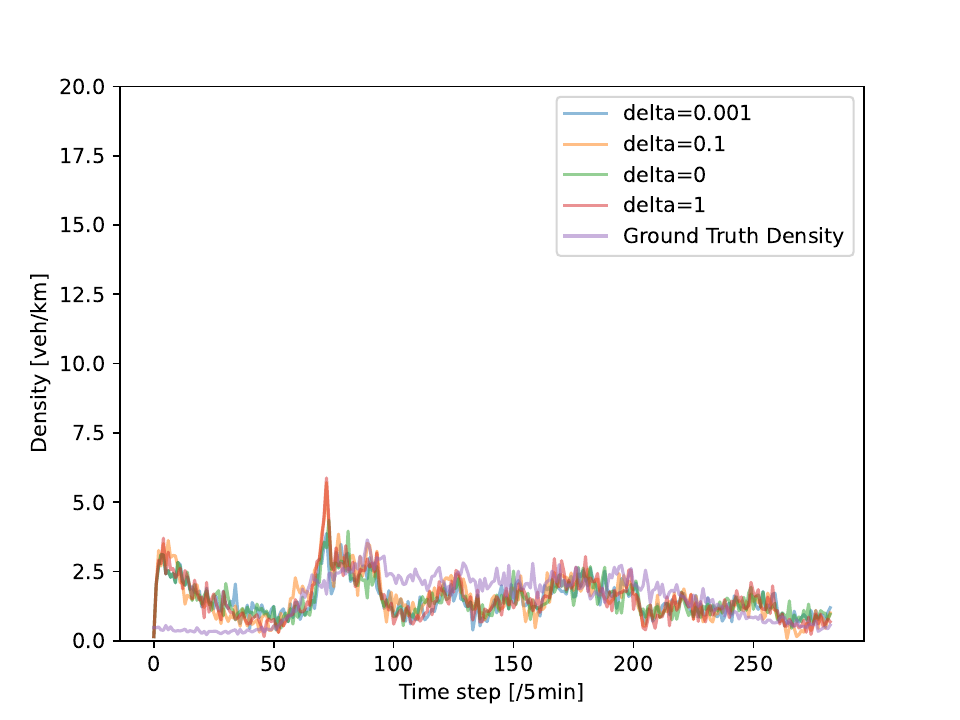}
    \caption{Location 5}
    \label{fig:Location_4_Oly_sensitivity}
\end{subfigure}%
\begin{subfigure}[b]{0.5\columnwidth}
    \centering
    \includegraphics[width=\textwidth]{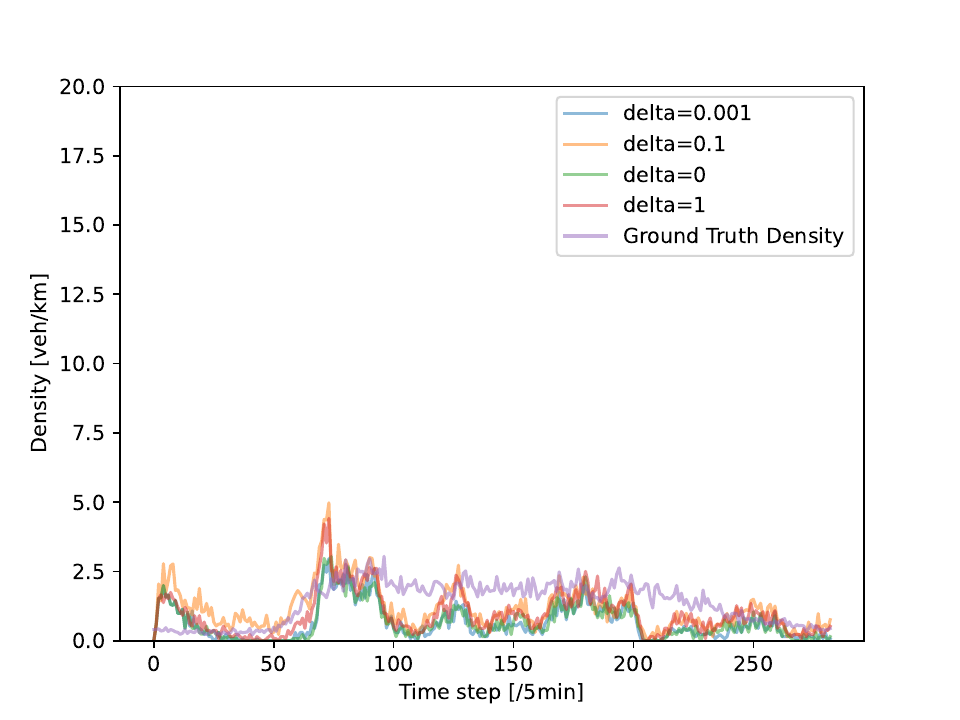}
    \caption{Location 6}
    \label{fig:Location_5_Oly_sensitivity}
\end{subfigure}
\caption{Sensitivity analysis with respect to $\delta$}
\label{fig:locations_oly_sensitivity}
\end{figure}

Fig.~\ref{fig:locations_oly_sensitivity} shows the sensitivity of the time series of traffic density at each location. When adjusting to new traffic patterns, $\delta$ has an important role for the learning time. Fig.~\ref{fig:Location_0_Oly_sensitivity}  illustrates that the model learns new traffic patterns fastest when $\delta=0.1$, and interestingly demonstrates that a lower $\delta$ does not always result in faster learning. The learning time for $\delta=0$ is even longer than that for $\delta=1$. This could be because inherent traffic demand patterns exist even in non-recurrent cases, and a priori demand information can improve learning time. Identifying the optimal $\delta$ to achieve the fastest learning time is an interesting direction for future work. Nevertheless, once the model adjusts to new traffic patterns, $\delta$ does not significantly impact the results, as shown in Table.~\ref{tab:mse_comparison_sensitivity}.

\begin{table}[h]
    \centering
        \caption{Sensitivity analysis with respect to $\delta$}
    \label{tab:mse_comparison_sensitivity}
    \begin{tabular}{ccccc}
        \textbf{Location} & \textbf{MSE ($\delta=0$)} & \textbf{MSE ($\delta=0.001$)} &\textbf{MSE ($\delta=0.1$)} &\textbf{MSE ($\delta=1$)}  \\
        \hline
        0 & 8.4 & 8.3 & 4.0 & 6.0 \\
        1 & 3.1 & 3.0 & 2.5 & 2.8 \\
        2 & 1.0 & 1.0 & 0.9 & 1.1 \\
        3 & 0.7 & 0.7 & 0.8 & 0.8 \\
        4 & 0.8 & 0.8 & 0.8 &  0.9 \\
        5 & 1.0 & 1.1 & 0.8 &  0.7 \\
        \hline
        &&&& [(veh/km)$^2$]
    \end{tabular}

\end{table}

\section{Conclusions}
This paper presents a novel hybrid neural network (NN) architecture for real-time origin-destination (OD) demand calibration. By leveraging real-time data and continuous NN parameter updates, our proposed framework can cope with both recurrent and non-recurrent traffic conditions. The integration of real-world and simulated traffic data allows for the real-time NN parameter updates by a proposed metamodel-based backpropagation, ensuring precise OD demand predictions  even under unforeseen traffic patterns. Additionally, the offline pre-training of the NN using the metamodel enhances computational efficiency.

Validation through a toy network and a case study of the Tokyo expressway corridor demonstrates the model's effectiveness in dynamically adjusting to changing traffic patterns in both recurrent and non-recurrent cases. The results highlight the potential of advanced machine learning techniques in developing proactive traffic management strategies via real-time simulation. 

There are several directions of future work. The first direction is extending the current framework from a single corridor to a more complex This can be done be done by integrating an appropriate deep learning technique, such as a graph neural network, along with a problem-specific metamodel \cite[e.g.,][]{osorio2019dynamic}.   Second, developing a comprehensive framework that leverages real-time simulations for traffic control, such as traffic signal control \citep{dantsuji2023perimeter, dantsuji2024hypercongestion} and congestion pricing \citep{zheng2016time} will be one of the interesting future directions. Another important direction is the development of a calibration framework that includes not only the OD demand but also the parameters of the traffic simulation models. This holistic approach will ensure that the entire simulation environment is accurately tuned to reflect real-world conditions, further enhancing the reliability and accuracy of traffic predictions.

\section*{Acknowledgement}
We thank the Tokyo Metropolitan government for providing the data. We also thank Eiji Hato for the valuable comments. 

\bibliographystyle{apalike} 
\bibliography{reference}
\end{document}